\documentclass{ieeeaccess}
\usepackage{amsmath,amssymb,amsfonts}
\usepackage{graphicx}
\usepackage{textcomp}
\usepackage{tabularx}
\usepackage{bytefield}
\usepackage{amsmath}
\usepackage{amsthm}
\usepackage{graphicx}
\usepackage{caption}
\usepackage{subcaption}
\usepackage{hyperref}
\usepackage{subcaption}
\usepackage{dblfloatfix} 
\graphicspath{{figures/}}

\def\BibTeX{{\rm B\kern-.05em{\sc i\kern-.025em b}\kern-.08em
    T\kern-.1667em\lower.7ex\hbox{E}\kern-.125emX}}
    
\begin{document}

\def\papertitle{Zero-Knowledge Proof of Traffic: A Deterministic and Privacy-Preserving Cross Verification Mechanism for Cooperative Perception Data}

\history{Date of publication xxxx 00, 0000, date of current version xxxx 00, 0000.}
\doi{10.1109/ACCESS.2023.0322000}

\title{\papertitle}
\author{
    \uppercase{Ye Tao}\authorrefmark{1}, \IEEEmembership{Student Member, IEEE},
    \uppercase{Ehsan Javanmardi}\authorrefmark{1}, \IEEEmembership{Member, IEEE},
    \uppercase{Pengfei Lin}\authorrefmark{1}, \IEEEmembership{Student Member, IEEE},
    \uppercase{Jin Nakazato}\authorrefmark{1}, \IEEEmembership{Member, IEEE},
    \uppercase{Yuze Jiang}\authorrefmark{1},
    \uppercase{Manabu Tsukada}\authorrefmark{1}, \IEEEmembership{Member, IEEE},
    \uppercase{Hiroshi Esaki}\authorrefmark{1}, \IEEEmembership{Member, IEEE}
}

\address[1]{
Graduate School of Information Science and Technology, The University of Tokyo, 1-1-1, Yayoi, Bunkyo-ku, Tokyo, 113-8657, Japan}
\tfootnote{
These research results were obtained from the commissioned research Grant number \#01101 by the National Institute of Information and Communications Technology (NICT), Japan. \\
This work was partly supported by the Japan Society for the Promotion of Science (JSPS) KAKENHI (grant number: 21H03423)
}

\markboth
{Tao \headeretal: Zero-Knowledge Proof of Traffic}
{Tao \headeretal: Zero-Knowledge Proof of Traffic}

\corresp{Corresponding author: Ye Tao (tydus@hongo.wide.ad.jp).}

\begin{abstract}
Cooperative perception is crucial for connected automated vehicles in intelligent transportation systems (ITSs); however, ensuring the authenticity of perception data remains a challenge as the vehicles cannot verify events that they do not witness independently.
Various studies have been conducted on establishing the authenticity of data, such as trust-based statistical methods and plausibility-based methods.
However, these methods are limited as they require prior knowledge such as previous sender behaviors or predefined rules to evaluate the authenticity.
To overcome this limitation, this study proposes a novel approach called zero-knowledge Proof of Traffic (zk-PoT), which involves 
generating cryptographic proofs to the traffic observations.
Multiple independent proofs regarding the same vehicle can be deterministically cross-verified by any receivers without relying on ground truth, probabilistic, or plausibility evaluations.
Additionally, no private information is compromised during the entire procedure.
A full on-board unit software stack that reflects the behavior of zk-PoT is implemented within a specifically designed simulator called Flowsim.
A comprehensive experimental analysis is then conducted using synthesized city-scale simulations, which demonstrates that zk-PoT's cross-verification ratio ranges between 80~\% to 96~\%, and 80~\% of the verification is achieved in 2 s, with a protocol overhead of approximately 25~\%.
Furthermore, the analyses of various attacks indicate that most of the attacks could be prevented, and some, such as collusion attacks, can be mitigated.
The proposed approach can be incorporated into existing works, including the European Telecommunications Standards Institute (ETSI) and the International Organization for Standardization (ISO) ITS standards, without disrupting the backward compatibility.
\end{abstract}
 

\titlepgskip=-21pt

\maketitle

\section{Introduction}\label{sec:intro}
\PARstart{R}{oad} transportation has been one of the most essential services for human mobility since ancient times.
It has undergone minimal changes despite the passage of time, with the driver of the vehicle still being responsible for determining its operation based on the surrounding environment.
Modern vehicles are fitted with various sensors, such as LiDAR~\cite{Roriz2022-gy}, millimeter wave~\cite{Martin-Vega2018-va}, and stereo cameras~\cite{Mao2023-zh}, which provide data to support driving assistance features and enable autonomous driving capabilities.
However, the effectiveness of these sensors is limited by their inherent locality, i.e. sensing ranges~\cite{Cui2022-tj}.
As sensors are attached to the vehicle, they present a view that is comparable to that available to the driver.
Consequently, they cannot detect objects that are beyond the sensing ranges or obstructed by obstacles.

\begin{figure}[t!]
  \centering
  \includegraphics[width=\linewidth]{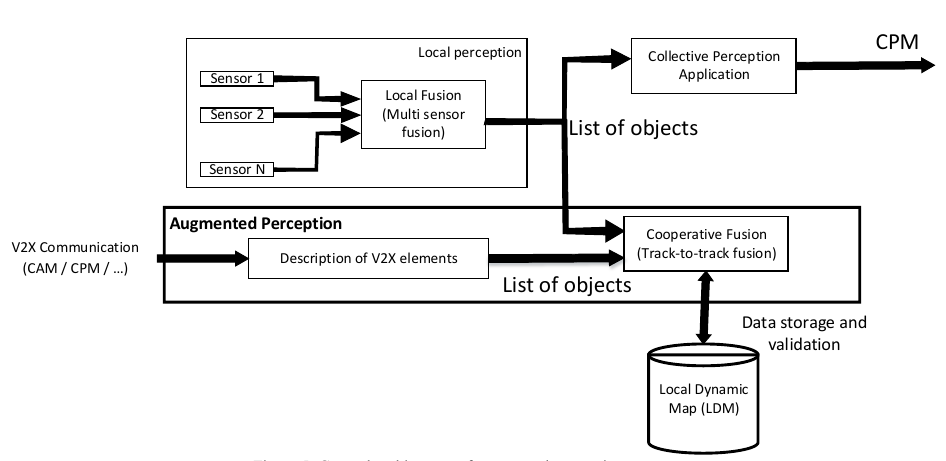}
  \caption{General structure of CPS}
  \label{fig:cps}
\end{figure}

Cooperative perception is a groundbreaking technology to break such limitations.
It can expand our field of vision and reduce blind spots by leveraging the sensors of other vehicles.
The standardization of cooperative perception in the International Organization for Standardization (ISO)~\cite{International_Standard_Organization2020-wf} and the European Telecommunications Standards Institute (ETSI) ~\cite{Etsi2023-vq} demonstrates its significance in the field.
The Intelligent Transportation System (ITS) station architecture standardized in the ETSI designates the collective perception service (CPS) as a crucial safety-related application protocol, as depicted in Fig.~\ref{fig:cps}.
The protocol utilizes collective perception messages (CPMs) to share information regarding the observed vehicles, such as the position, velocity, and other important information with other vehicles.
Although the CPS standard has been published recently, multiple open-source and proprietary CPM implementations have already been published~\cite{Asabe2023-hk}, which greatly facilitates further research and development.

The CPMs received from other vehicles can influence the decisions made by the vehicles.
Thus, ensuring the authenticity of these messages is crucial for road safety, as the CPMs are generated by vehicles rather than a centralized and trusted authority~\cite{Hasrouny2017-xb,Golle2004-px}. 
Unlike a centralized authority, individual vehicles may be incentivized to deceive other vehicles to maximize their own profits~\cite{Leinmuller2008-uw}.
As an example, a malicious vehicle could simulate congestion on the road ahead by sending synthesized CPMs, prompting other vehicles to take an alternative route, to clear the way for themselves.

Numerous solutions have been suggested to detect and eliminate such fraudulent activities.
One such solution is the Public Key Infrastructure (PKI), which has been standardized in the VANET~\cite{Etsi2022-xf}.
This standard mandates that every message transmitted over the VANET must be signed using a secret key provided by law enforcement agencies (LEAs).
Consequently, every node in the network is forced to obtain the keys from LEAs before sending any message.

Nevertheless, vehicles can still send fraudulent messages on purpose and sign them with their valid LEA-issued keys.
Most existing security solutions for the Internet are impractical owing to VANET's highly decentralized and mobile nature.
Various trust management methods have been proposed to overcome this problem~\cite{Minhas2011-bw, Raya2008-ja, Yang2019-oj, Suo2019-no}.

In this work, we proposed a zero-knowledge proof (ZKP) based deterministic traffic cross-verification method called zero-knowledge Proof of Traffic (zk-PoT).
Zk-PoT enables vehicles to prove the existence of vehicles they observed independently, then enables the remote parties to deterministically cross-verify the observations without any knowledge about the ground truth.
Subject to the zero-knowledge property, the public cannot gain any extra information in the whole process, thus the privacy of the observed vehicles is preserved.


The proposed mechanism can enhance the security, efficiency, and verification latency while preserving the privacy of the target vehicles.
Integrating the zk-PoT into existing cooperative perception standards such as ISO and ETSI requires minimal changes to the architecture, while still maintaining backward compatibility.
Additionally, it can be used as a bootstrap method alongside existing trust management methods.

The basic concept of this work is proposed in our published paper~\cite{Tao2023-nn}.
In this paper, we refine the original concept and propose a more concrete design and a comprehensive quantitative analysis.
Furthermore, a detailed proof-of-concept implementation is performed for more realistic evaluations.
As the ETSI ITS CPS standard, which is considered the basis of this work, is updated according to the recently published version of ETSI TS 103 324~\cite{Etsi2023-vq}, the zk-PoT is further adapted to the latest version.
Additional results regarding efficiency, security, and communication overhead are analyzed based on the new implementation.

The rest of this paper is organized as follows.
Section~\ref{sec:related} reviews the related works and highlights the challenge of balancing location privacy and trust management.
We also discuss the cryptographic tools required for the proposed solution, including the Elliptic Curve Digital Signature Algorithm (ECDSA) and zero-knowledge proofs.
Section~\ref{sec:problem} defines the problem, outlines reasonable assumptions, and presents the approaches to address the problem.
Section~\ref{sec:proposal} describes how the problem can be transformed into a cryptographic problem and then be solved using zero-knowledge proofs, and the means of applying the proposed solution by extending the existing CPS standard.
Section~\ref{sec:evaluation}, presents quantitative analyses performed using the previously proposed simulator, Flowsim~\cite{Tao2023-oe}.
Section~\ref{sec:analysis}, analyzes the robustness of the proposed method against common threats including various attacks and privacy leakage.
Lastly, Section~\ref{sec:conclusion} summarizes the contributions of the proposed method and presents future research directions.

\section{Related Work}
\label{sec:related}

\subsection{Misbehavior Detection}

Misbehavior detection (MBD) is the process of identifying and mitigating malicious or inappropriate behavior within VANETs.
This involves identifying vehicles or nodes that engage in unauthorized or harmful activities, such as sending false data or participating in attacks.
MBD methods can be categorized into four types by two orthogonal criteria: node-centric vs. data-centric, autonomous vs. collaborative~\cite{Xu2022-ks}.
Node-centric methods perform evaluations based on the behavior of individual vehicles, whereas data-centric methods analyze the content and characteristics of the transmitted data.
Local MBD methods involve independent node detection without depending on other nodes, whereas collaborative MBD methods employ neighbor nodes to verify the data and identify misinformation.
Zacharias et al.~\cite{Zacharias2018-ao} proposed an autonomous node-based MBD system based on the local traffic density.
This approach utilized multiple independent sensors to measure the traffic density and combined the evidence using the Dempster rule of combination to detect misbehavior, particularly focusing on illusion attacks.
Al-Ali et al.~\cite{Al-Ali2020-ve} proposed a blockchain-based collaborative approach to validate traffic events and authenticate vehicles in VANET.
It utilizes the reputation scores, proof of authority (PoA) and proof of event (PoE) consensus algorithms, as well as mutual authentication between vehicles and roadside units (RSUs) to improve the event validation accuracy and detect internal attackers.

Numerous data-centric methods have also been proposed over the years.
Ghosh et al.~\cite{Ghosh2010-dv} proposed an autonomous MBD scheme for the post crash notification (PCN) application.
It involves identifying the root cause of misbehaviors by constructing a cause tree and using logical reduction.
This scheme achieves adequate detection rates and exhibits robustness to small errors in the parameter estimation.
Lo et al.~\cite{Lo2007-he} introduced a security threat in VANETs called the illusion attack, where an adversary broadcasts false traffic warning messages based on the current road conditions, thereby creating an illusion for nearby vehicles.
This illusion can manipulate the drivers' behaviors, leading to car accidents, traffic jams, and decreased VANET performance.
The authors proposed an autonomous data-centric security model called the plausibility validation network (PVN) to address this issue by cross-verifying the plausibility of incoming message fields.
The problem they were aiming to solve is very similar to our problem, as we also utilize cross-verification, due to which the requirements are very similar.
Ercan et al.~\cite{Ercan2022-hi} proposed a machine learning-based Intrusion Detection System (IDS) to detect position falsification attacks in VANETs.
The IDS utilizes three new features corresponding to the sender's position, along with the k-nearest neighbor (kNN) and random forest (RF) classification algorithms.
The results demonstrate that the proposed mechanism outperforms the existing approaches in terms of classification performance and computation time.
Kristianto et al.\cite{Kristianto2023-fw} proposed a semi-supervised federated learning MBD system for V2X communications.
This model addresses the challenges of limited labeled data and bandwidth consumption by leveraging semi-supervised learning and federated learning approaches.
The experimental results show that the model achieves high performance, outperforming centralized supervised learning methods regarding the F1-score, recall, and reduced bandwidth utilization.

Although all the above works are proposed, the MBD problem still cannot be solved perfectly in all situations.
Data-centric methods require a significant amount of data, presenting higher channel occupation, packet loss, and increased latency.
Furthermore, feasibility evaluations are the only options available in cases where the ground truth data are missing or difficult to obtain.
In entity-centric methods, data correctness is still a major problem owing to the presence of attackers and sensor limitations.

\subsection{The Dilemma of Location Privacy and Trust Management}

Connected autonomous vehicles (CAVs) are designed to share their position and velocity information with other CAVs on the road, unlike conventional vehicles that share either limited or no information.
However, preserving location privacy while ensuring accurate trust evaluation is challenging for most trust management methods as they are required to track the vehicles' historical behaviors.

Efforts were made to balance the dilemma, some by introducing a pseudonymous authentication scheme that can be used to protect the vehicles' location privacy~\cite{Mansour2018-tz}; however, this increases the difficulty of tracking specific vehicles for trust management.
Recently, some methods that employ modern cryptographic tools such as self-blindable signatures and zero-knowledge proofs were also proposed for specific applications~\cite{AlMarshoud2022-xx}.

\subsection{Digital Signature and ECDSA}
A digital signature is a mathematical scheme that uses asymmetric cryptography to verify the authenticity of digital messages or documents.
The Digital Signature Algorithm (DSA) is a standard of digital signature based on the discrete logarithm problem.
ECDSA~\cite{Johnson2001-sx} is a variant of the DSA that employs elliptic curve cryptography, which is widely used in VANET and V2X applications, providing authentication, integrity protection, and privacy enhancements~\cite{Hammi2022-fg,Sedar2023-fu,Kushwah2019-tz}.
ECDSA also serves essential roles in the existing ETSI ITS standards~\cite{Etsi2022-xf,Etsi2015-im}.
Many experiments and analyses regarding ECDSA's performance and impact on V2X communication are carried out~\cite{Cirne2019-gn,Hakeem2020-md,Pollicino2020-ow,Petit2009-fo,Rufino2019-ze}, providing a comprehensive understanding about ECDSA's performance in the context of ITS and VANETs.

\subsection{Zero-Knowledge Proofs}
In 1989, Goldwasser et al. proposed the zero-knowledge proof (ZKP) system~\cite{Goldwasser1989-js}.
This system allows one party, called the prover, to convince another party, called the verifier, that a specified statement is true, without revealing any additional information to the verifier, except for the truth of the statement.
While ZKP systems are widely used in cryptocurrency protocols such as Zerocoin~\cite{Miers2013-za} and Zerocash~\cite{Ben_Sasson2014-ka}, they are still relatively new in other fields of study.

Nowadays, research based on zero-knowledge proofs is emerging in the ITS field.

McEntyre et al.~\cite{McEntyre2022-ur} present a privacy-preserving electronic toll collection (ETC) protocol for V2X communications, utilizing a ZKP challenge set to ensure security while addressing the limitations of embedded technology.
The protocol achieves toll verification without disclosing sensitive subscriber information, such as GPS location, by generating unique randomized challenges localized to toll areas, thereby reducing the likelihood of false tolling.

Chaudhry et al.~\cite{Chaudhry2021-yr} introduce a secure communication framework for vehicle-to-healthcare everything (V2HX) communications in fog computing environments, utilizing a combination of ZKP and statistical fingerprinting (SF) protocols.
The proposed framework enables vehicle authentication through ZKP and ensures secure communication between VANETs and healthcare enterprises through SF.

Rasheed et al.~\cite{Rasheed2020-aw} present an Adaptive Group-based Zero Knowledge Proof Authentication Protocol (AGZKP-AP) for VANETs, addressing privacy concerns in authentication.
The protocol enables anonymous authentication, distributed privilege control, and customizable privacy settings for users while minimizing the disclosure of authentication parameters, enhancing user privacy in VANETs.

Li et al.~\cite{Li2023-qu} introduce an aggregated zero-knowledge proof and blockchain-based authentication system for privacy-preserving identity verification in autonomous truck platooning, addressing security and privacy concerns.
The proposed approach enhances security, provides fast performance, and ensures data integrity while allowing truck companies to define access control policies.
Experimental results on the Hyperledger platform demonstrate the system's feasibility for real-world truck platooning applications.

\section{Problem Statement, Assumptions and Approaches}
\label{sec:problem}
\subsection{Problem Statement}

\begin{figure*}[tb]
    \centering
    \includegraphics[width=.85\linewidth]{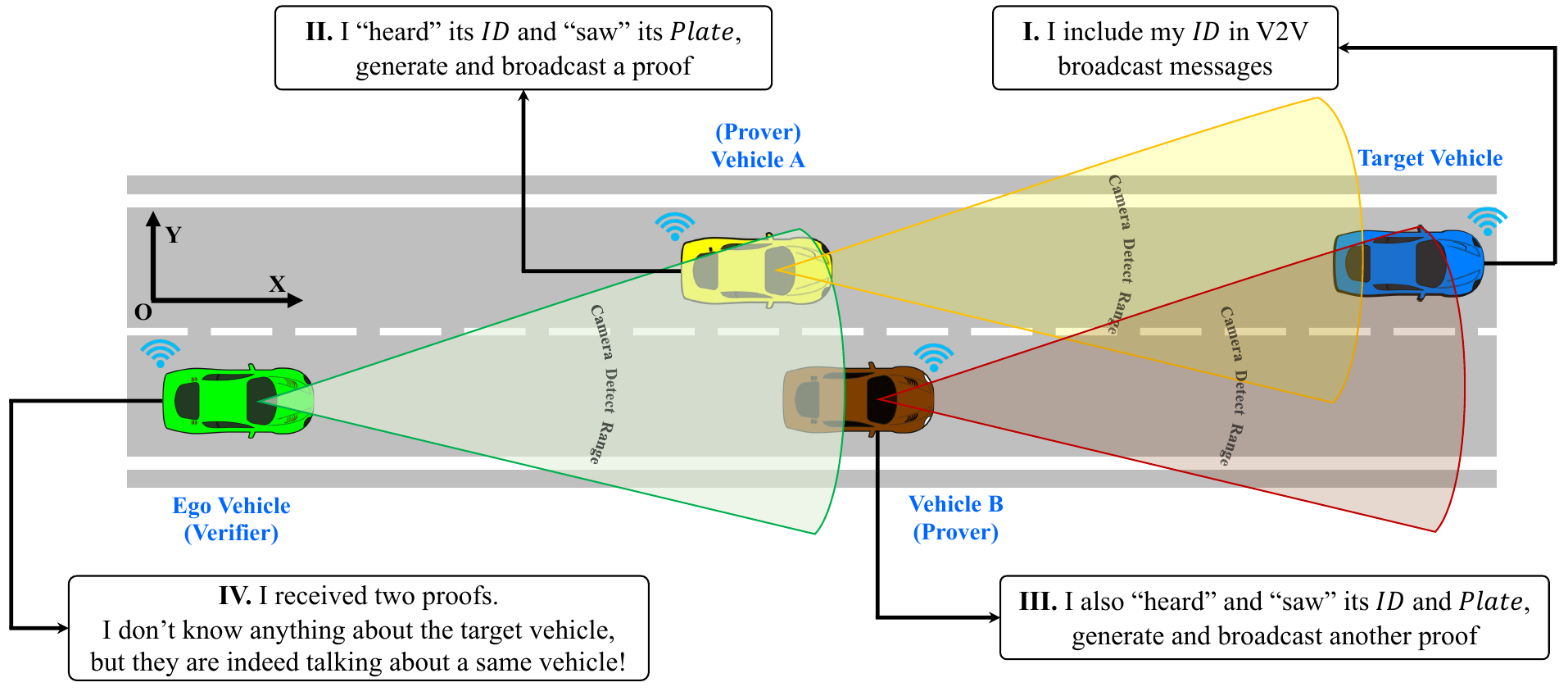}
    \caption{Procedure of a Zero-Knowledge Proof of Traffic}
    \label{fig:scenario}
\end{figure*}

The cooperative perception ability provided by CPMs is significant; however, the potential risks associated with such messages must be considered.
Blindly trusting CPMs could lead to suboptimal or even risky vehicle decisions, which can considerably compromise road safety.

Some studies~\cite{Minhas2011-bw} are based on trust estimation and management methods that depend on past statistics, making it nearly impossible to detect one-shot misinformation.
Other studies rely on assessing the ``plausibility'' of data, which also necessitates the use of past statistics.
Furthermore, these models cannot deterministically verify data as they are statistical models.

Therefore, we designed a cryptography-based mechanism that allows vehicles to prove their traffic observations, enabling other vehicles to cross-verify and \textit{deterministically} trust the observations without knowing the ground truth.

Our study is primarily focused on proving the existence of observed vehicles as ``existence'' is the most fundamental and accurate event that occurs on the road, which remains unchanged despite the high speed of vehicles.
This is contrary to a vehicle's location, which can change quickly, leading to inaccuracies.
Although fundamental, proving existence can already block many data-fabricating attacks, such as the most widely discussed phantom vehicle attack.

\subsection{Assumptions}
In this study, we make several assumptions.
Firstly, We assume that all vehicles have a front camera system and can recognize the other vehicles' number plates by computer vision.
We also assume that every vehicle on the road joins VANET, enabling all vehicles to perceive and share information with other vehicles.
Consequently, we do not differentiate between conventional vehicles vs. vehicles with cameras, connected vs. non-connected vehicles, in our subsequent discussions.

Additionally, we assume that a public key infrastructure (PKI) run by LEAs is deployed to all the infrastructures and vehicles.
Furthermore, vehicle certificates with key pairs are distributed to all the vehicles by LEAs; each vehicle signs every message it sends and uses the public key to verify the integrity and authority of the messages they receive.
We also assume that all the vehicle certificates are provided by a pseudonym system with perfect unlinkability.
This implies that after a vehicle changes its pseudonym, i.e. its pseudonymous certificate, the new pseudonym cannot be linked with the old one.

Based on these assumptions, we define ``A heard B'' as A receiving a message that contains B's pseudonym from V2X communication.
Conversely, we define ``A saw B'' as A identifying B's number plate, extracted from the video feed of its front camera.

\subsection{Approaches}
We can design a na\"ive ``proof system'', which lets vehicles broadcast the value of vehicle number plates that they observe.
However, such a ``proof system'' is not secured as the plate numbers can be easily synthesized, which makes it subject to data forging, replay attacks, and other threats.
Moreover, the location privacy of the observed vehicles is compromised, since the number plates are in plaintext and can be used to track vehicles by malicious parties.

To address these issues, we propose a zero-knowledge proof (ZKP) based system that enables vehicles to prove that they have observed another vehicle.
Based on the definition of ``seeing'' and ``hearing,'' we consider that the vehicle that can link the number plate and pseudonym must have a close observation of the target vehicle, presenting strong evidence of the existence of the target.
Additionally, our ZKP-based system ensures that both the pseudonyms and number plates remain undisclosed in the proofs, thus ensuring location privacy.

In our system, it is difficult if not impossible to falsify the existence of a random vehicle.
To overcome this limitation, we propose a novel approach that employs a cross-verification scheme.
Our approach involves multiple vehicles generating individual proofs for the same target vehicle.
Thus, any third party can verify the existence of the target vehicle by comparing multiple proofs without requiring any information regarding the specific vehicle.

\section{Zero-Knowledge Proof of Traffic}
\label{sec:proposal}
In this section, we present our proposed solution, Zero-Knowledge Proof of Traffic (zk-PoT), to the problem of verifying whether two vehicles have observed the same target vehicle while preserving their location privacy.
We convert the problem into a cryptographic protocol called zero-knowledge proof of shared secret (zk-PoSS).
We then provide a solution to this model and apply it to the existing ETSI standard by extending the packet structure and station behaviors.
We use ETSI standards as an example; our approach can also be applied to other cooperative perception systems, such as ISO standards.

\subsection{Problem Conversion}
Fig.~\ref{fig:scenario} depicts a scenario where vehicles A and B observe a vehicle named Target and acquire its pseudonym, $ID$, and number plate, $Plate$.
These observations essentially yield a~\textit{shared secret}, which is derived from the target vehicle's $ID$ and $Plate$.
To prove that they indeed know the shared secret, without disclosing the plaintext $ID$ and $Plate$, we proposed a cryptographic protocol called zk-PoSS.
This model involves creating two distinct proofs using a one-way proof function, $F_{proof}$, using a shared secret, $SS$, and two cryptography salts, $m$ and $m^\prime$.
Subsequently, another function, $F_{verify}$, is used to determine if the pairs, $(P, m)$, and $(P^\prime, m^\prime)$, are correspond to the same shared secret, $SS$.
\begin{figure*}[t!]
    \centering
    \includegraphics[width=.9\linewidth]{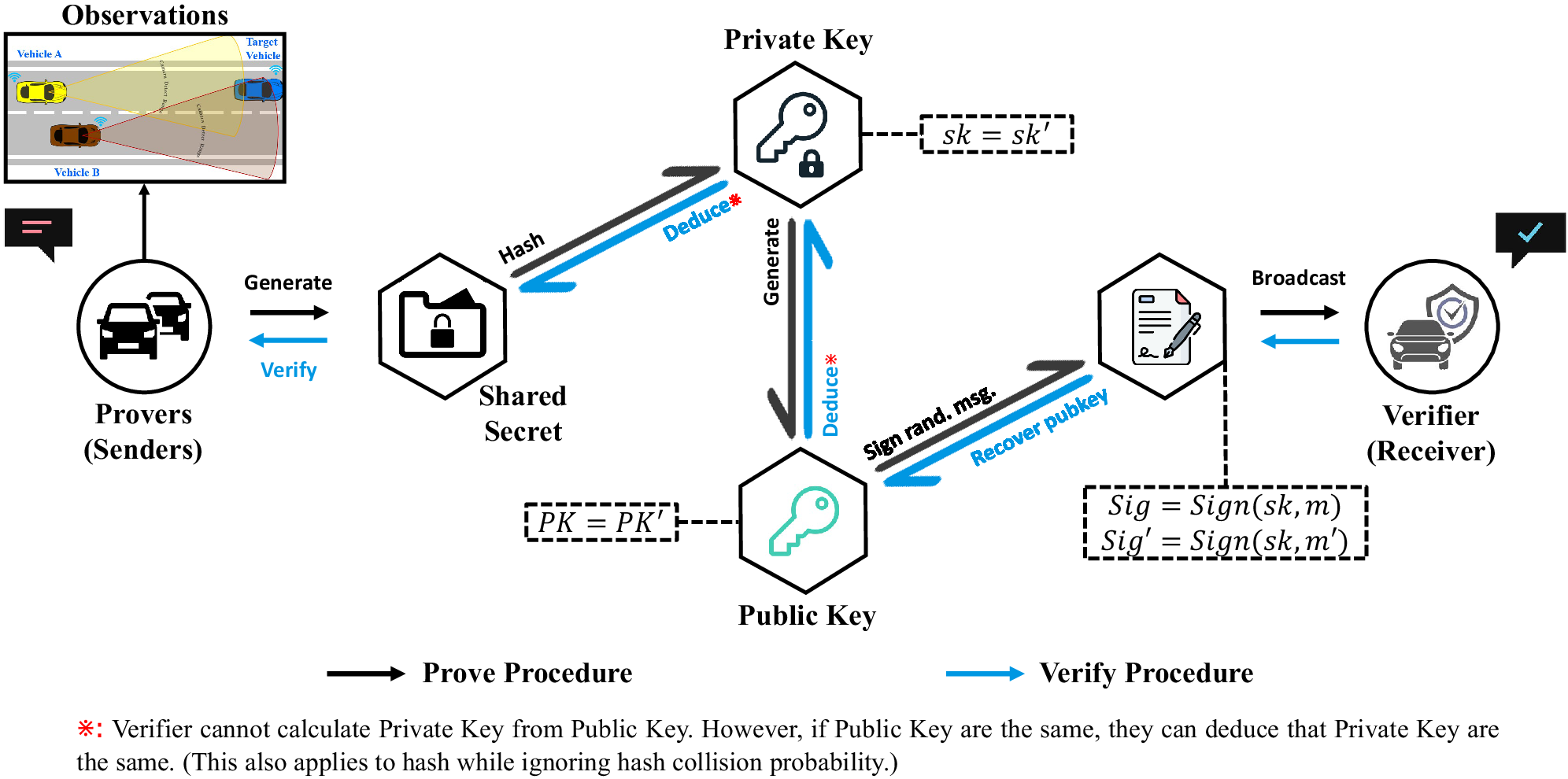}
\caption{Overview of Zero-knowledge Proof System}
    \label{fig:overview}
\end{figure*}

By doing so, the problem of proving a common observation to a vehicle could be transferred to the cryptographic model of proving the shared secret:

\begin{align}
\begin{split}
    &P = F_{proof}(SS, m) \\
    &P^\prime = F_{proof}(SS, m^\prime) \\
    &P \neq P^\prime \\
    &F_{verify}(P, m, P^\prime, m^\prime) = True
\end{split}
\end{align}

\subsection{Zero-Knowledge Proof of Shared Secret}
After converting the practical problem of zk-PoT into the theoretical problem zk-PoSS, it can be solved using the conventional ECDSA.

The conceptual overview of zk-PoSS is shown in Fig.~\ref{fig:overview}.
Concatenating the ID and plate of a specific observed vehicle, the provers essentially possess the same shared secret with sufficient entropy.
Since the private key of ECDSA is just a plain integer, they can interpret the hash of such a shared secret as a private key.
Therefore, two valid but different signatures become proof that the provers possess the same private key.
In the proposed scheme, holding the private key implies knowledge of the shared secret.

Utilizing this feature of digital signature, two parties can create different proofs regarding the shared secret, by signing different random messages as salts.
Subsequently, a third party that received both proofs can recover the public keys from the signatures, and verify if the two public keys are identical.
If these public keys are identical, it can deduce that the private keys are identical; and that the private key is, in essence, a hash of the shared secret.
Therefore, from the proof by contradiction, the only feasible option is that both indeed share the same secret.

The exact protocol is defined in Protocol \textbf{zk-PoSS}.

\hfill \\

\noindent Protocol:\textbf{zk-PoSS}

\noindent\textbf{Preconditions: Common Agreement} \\
\indent Prior to initiating the protocol, there is a set of common parameters that all parties must agree upon.
    This ensures that all parties are working with the same fundamental assumptions, enabling coherent and meaningful communication.
    \begin{enumerate}
        \item All parties agree with the same set of ECDSA parameters: $$\{CURVE, G, n\}$$
        These parameters define the elliptic curve, the base point, and the order of the base point, respectively.
        \item All parties adopt the same cryptographic hash function, denoted by $H$.
        This function is capable of producing a fixed-length output that matches the binary length of $n$. This uniformity ensures consistency across all hashed outputs.
    \end{enumerate}

\noindent\textbf{Prover: Proof Construction} \\
\indent The prover's role is to generate a proof of knowledge to a secret, without revealing the secret itself. The following steps outline this process.
    \begin{enumerate}
        \item The prover computes the hash of the secret $SS$ using the agreed cryptographic hash function: $$sk = H(SS)$$
        Here, the hashed secret, $sk$, could serve as a private key in the ECDSA context.
        \item There may be rare cases where $sk >= n$ since in ECDSA $n$ is a prime which is slightly less than a power of two.
        To address these outliers, we perform an iterative process of recalculating the hash by repeating $SS$, until $sk < n$.
        \item The prover then computes $$PK = G \times sk$$ on curve $CURVE$.
        The generated pair ($sk$, $PK$) can be interpreted as the private key and public key of the ECDSA algorithm, respectively.
        \item Using the private key $sk$, the prover signs a random message $m$ with the function: $$Sig = ECDSA\_Sign(sk, m)$$ This signature attests to the ownership of $sk$.
        \item The prover publishes the message $m$ and its corresponding signature $Sig$, available for any verifier to check.
    \end{enumerate}

\noindent\textbf{Verifier: Pairing and Verifying Proofs} \\
\indent The verifier's role is to match and validate the proofs provided by the provers, thus affirming the provers' common knowledge of the shared secret. Here are the steps a verifier must follow.
    \begin{enumerate}
        \item The verifier recovers the public key $PK$ from the message $m$ and signature $Sig$ using the function: $$PK = ECDSA\_Recover(m, Sig)$$
        The recovered public key is then stored in a database locally.
        \item If the database already contains another message $m'$ with corrsponding valid signature $Sig'$ which is signed using the same public key $PK$, it implies that both provers of $m$ and $m'$ possessing such $PK$ share the same secret.
        This is a significant conclusion as it allows us to infer shared knowledge without revealing the secret itself.
    \end{enumerate}

\hfill \\

A zero-knowledge proof system must satisfy three properties~\cite{Goldwasser1989-js}:
\begin{enumerate}
    \item \textit{Completeness}: If the statement is true, the verifier can be convinced by the prover.
    \item \textit{Soundness}: If the statement is false, the verifier cannot be convinced by the prover, even if the prover tries to cheat.
    \item \textit{Zero-knowledge}: The verifier cannot gain any knowledge apart from the truth of the statement.
\end{enumerate}
Zk-PoSS is considered a zero-knowledge proof system because it satisfies all three properties.
\begin{enumerate}
    \item \textit{Completeness}: If $PK$s are identical, the verifier could deduce that the $SS$ used to generate the $PK$s are identical, although it cannot know the actual secret.

    \item \textit{Soundness}: If $SS$ are not identical, then the same $sk$ can only be produced through a hash collision, which is a computationally difficult task. Thus, a cheating prover has a negligible chance of convincing the verifier.

    \item \textit{Zero-Knowledge}: In the whole process, the verifier cannot produce any information regarding $sk$ or $SS$ from $PK$s. This property is guaranteed by any public key cryptography, including ECDSA.
    
\end{enumerate}

\begin{figure*}[htb]
    \centering
    \includegraphics[width=.9\linewidth]{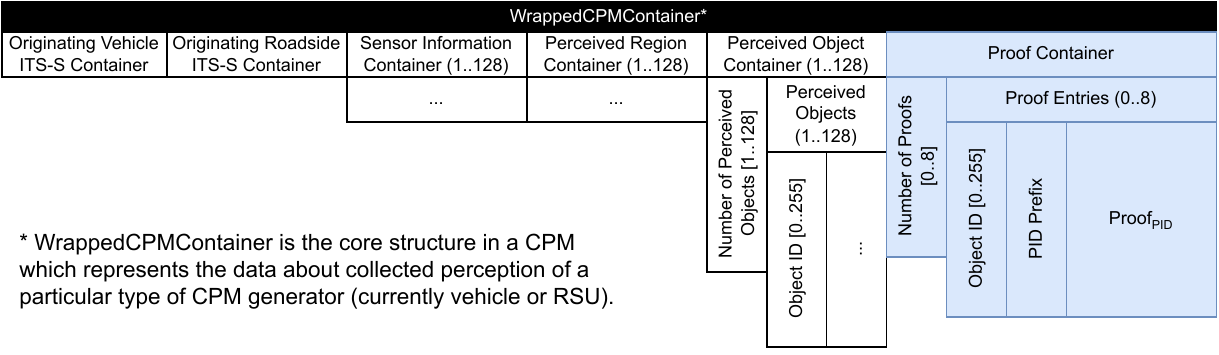}
    \caption{Structure of the extended part of collective perception message (CPM) for zk-PoT}
    \label{fig:cpm-extended}
\end{figure*}

\subsection{Integrating zk-PoT into the ETSI CPS standards}
The ETSI CPS system can be updated by incorporating the zk-PoSS protocol to provide proof of traffic capability.
The standard curve \texttt{secp256k1} is used with a 256-bit key length, which will produce a 65-byte long signature.
The message, $m$, used in zk-PoSS is the prover's pseudonym, which is selected for its high entropy and frequent transmission in every packet, enabling bandwidth conservation.
Furthermore, it helps in preventing the na\"ive replay attack, which will be discussed in Section~\ref{sec:analysis}.

\subsubsection{Changes to the CPM packet structure}

The \texttt{WrappedCPMContainer} structure in the CPM has been extended with an additional section called the \texttt{Proof Container}, which is located at the end, contains a list of 0 to 8 \texttt{Proof Entries}, as shown in Fig.~\ref{fig:cpm-extended}.
Each \texttt{Proof Entry} contains an \texttt{object ID} that is present in the perceived object container, a 32-bit prefix of the vehicle's ID to assist quick filtering, along with the actual proof of this ID.

The ASN.1 definition of \texttt{ProofEntry} is as follows.

\begin{verbatim}
ProofEntry ::= SEQUENCE{
    objectID Identifier2B,
    pidPrefix Integer32,
    v BOOLEAN,
    r OCTET STRING (SIZE(32)),
    s OCTET STRING (SIZE(32))
}
\end{verbatim}

The \texttt{objectID} field represents the ID of the target vehicle assigned in \texttt{PerceivedObjects},
the \texttt{pidPrefix} field represents the 32-bit prefix of the target vehicle's pseudonym,
and the three fields \texttt{V}, \texttt{R}, and \texttt{S} represent the corresponding fields in the ECDSA signatures, as defined in \cite{Brown2009-bu}.

It is recommended that the proofs be sent intermittently rather than in every CPM to minimize the amount of data transmitted and preserve bandwidth,
As the CPM can be broadcasted at a high frequency of up to 10 Hz~\cite{Etsi2019-mg}, the length of the proofs (71 bytes per \texttt{proof entry} under ASN.1 UPER encoding) can add up quickly.
Therefore, it is encouraged to implement an inclusion management process to limit the number of \texttt{proof entry} in the prover; and the receiver should store each entry for a specified period.
The exact time interval for transmitting the proofs is not yet determined; however, sending them once every 3 s is considered sufficient to handle vehicle topology changes.

\subsubsection{Changes to the ITS Station}

\begin{figure}[b]
    \centering
    \begin{subfigure}[b]{\linewidth}
        \centering
        \includegraphics[width=\textwidth]{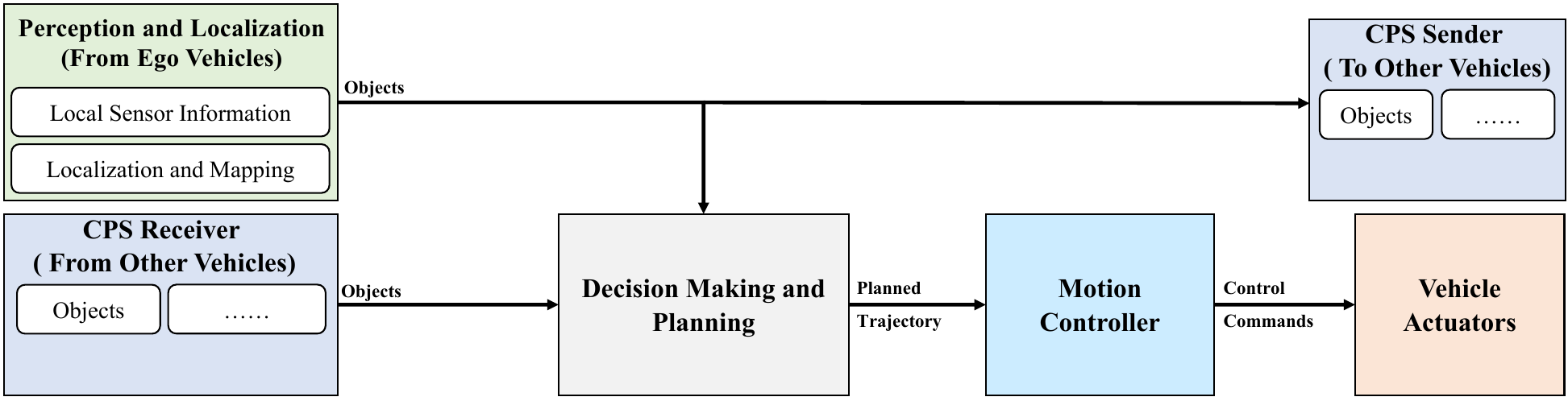}
        \caption{Original CPS}
        \label{fig:cps-original}
    \end{subfigure}
    \hfill \\
    \begin{subfigure}[b]{\linewidth}
        \centering
        \includegraphics[width=\textwidth]{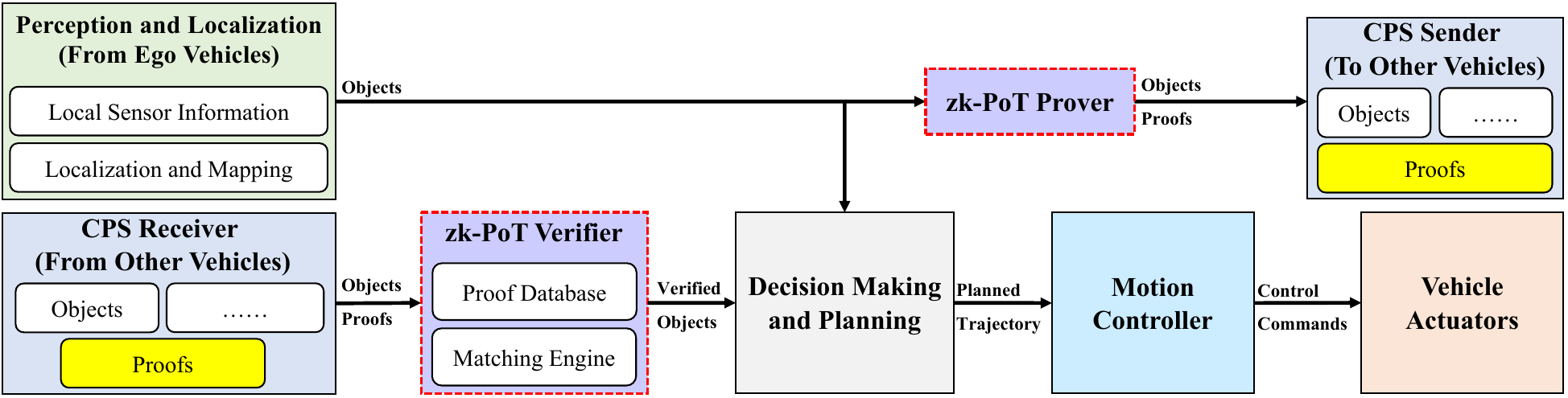}
        \caption{Extended CPS}
        \label{fig:cps-extended}
    \end{subfigure}
    \caption{Block Diagram of Original v.s. Extended CPS}
    \label{fig:cps-original-extended}
\end{figure}

Various implementations may produce different module designs; however, the fundamental structure of a vehicle equipped with the CPS should be similar to the one depicted in Fig.~\ref{fig:cps-original}.
The \texttt{receiver} module retrieves objects from the received CPMs, and in conjunction with locally perceived objects, feeds them into the \texttt{planning} module for further processing.
Additionally, the objects obtained from the \texttt{local} module are sent into the \texttt{sender} module to be packed as CPMs and broadcasted to the vehicles nearby.

\begin{table*}[ht]
    \centering
    \caption{Implemented vehicle types in Flowsim}
    \label{tab:flowsim-vehicle-types}
    \begin{tabular}{rl}
        \hline
        Name & Description \\
        \hline
        UnconnectedVehicle & Traditional vehicles with perception, but not connected to V2X \\
        ConnectedVehicle   & Perception and V2X communication capable vehicles, could understand CPMs \\
        PoTVehicle         & Proof of Traffic capable vehicles, could send and verify proofs \\
        SpamAttacker       & Attacker vehicles sending random fake objects to make the road look congested or blocking signals \\
        ReplayAttacker     & Attackers replaying CPM of both received and local perceived objects to confuse or overload others\\
        SilenceAttacker    & Selfish vehicles that only listen to V2V communications but do not contribute anything \\
        \hline
    \end{tabular}
\end{table*}
To accommodate the zk-PoT, we introduced two new modules: \texttt{prover} and \texttt{verifier}.
The \texttt{prover} module obtains objects from \texttt{local}, creates proof for each potential object, and then transfers the proofs along with the original objects to the \texttt{sender} module for CPM preparation and transmission, as shown in Fig.~\ref{fig:cps-extended}.
On the receiver side, the \texttt{verifier} module, positioned after the \texttt{receiver}, functions as a gatekeeper, ensuring that only verified received objects can be used for the \texttt{planning} module.
It receives raw objects and proofs, stores them in an internal database, and continually verifies these proofs.
When a match is found, the corresponding objects are considered authentic, and all the recent data of this object, including that stashed in the internal database, are passed to the \texttt{planning} module.
Additionally, any subsequent data from the same objects are immediately forwarded to the \texttt{planning} module.

Such design ensures minimal modifications to the pre-existing modules while incorporating the proof of traffic functionality.
Furthermore, this approach maintains backward compatibility with the current standards.

\section{Evaluation}
\label{sec:evaluation}
Although certain aspects, namely the vehicle traces, number plate recognition, and packet delivery, may not considered realistic enough within a simulated setting,
zk-PoT, functioning as a behavioral model, remains contingent solely upon the outcomes of these modules and is thereby not directly influenced by their simulated performance.
Thus, we consider that a simulation environment incorporating deterministic inputs suffices for the evaluation of a deterministic model.

Conversely, the pivotal consideration lies in evaluating the model's scalability across extensive geographic areas, ensuring its effectiveness in diverse traffic scenarios, including distinctions between highway and urban settings, as well as variations between rush hours and standard conditions.
Considering the cost and complexity of conducting a real-world large-scale experiment, a simulation-based approach is considered more viable.

To evaluate the efficiency and performance of the zk-PoT compared to non-verified CPS in a large-scale and realistic environment, we proposed the Flowsim~\cite{Tao2023-oe}, a modular simulation platform for microscopic behavior analysis of city-scale connected autonomous vehicles.
The Flowsim simulator has a simplified perception model, enabling us to evaluate the performance of zk-PoT in different visibility caused by diverse vehicle densities and occlusions.
Various microscopic simulations were conducted on the Flowsim to evaluate the robustness, attack resistance, and network efficiency of zk-PoT.

The vehicles currently implemented are listed in Table.~\ref{tab:flowsim-vehicle-types}.

%
%
%
%
%
%

\subsection{Parameter and metric settings}
A set of four experiments is designed to compare the zk-PoT with the conventional methods under different conditions:
\begin{itemize}
    \item Local Perception Only: all the vehicles operate standalone and do not communicate with each other.
    \item Conventional CPS: all the vehicles participate in the V2V communication and share their perception based on the ETSI CPS standards.
    \item Proof of Traffic (1s repeat): all the vehicles are capable of Proof of Traffic, which includes (up to 8) proofs for recently seen vehicles in the CPM they sent every second.
    \item Proof of Traffic (3s repeat): identical to PoT, except the vehicles only repeat proofs of the same vehicle once every three seconds. This setting can potentially decrease the overall bandwidth overhead while maintaining comparable performance.
\end{itemize}

These experiments are conducted in scenarios with different scales. Table.~\ref{tab:para} lists the common parameters throughout all the experiments.

\begin{table}[htb]
    \centering
    \caption{Common Parameter Settings}
    \label{tab:para}
    \begin{tabular}{ccc}
        \hline
        Parameter & Value \\
        \hline
        Vehicle Length & 4.0 meter\\
        Vehicle Width & 1.8 meter\\
        Numberplate Width & 0.35 meter\\
        Perception Distance & 65.0 meter\\
        Expected pseudonym valid time & 12 hours \\
        Pseudonym changing probability & 1/43200 per second \\
        Camera Sensing Angle & 120.0 degree\\
        Communication Range & 300 meter\\ 
        Communication Delay & 1 ms\\
        Packet Delivery Ratio & 80 \%\\
        \hline
    \end{tabular}
\end{table}

We collect the following metrics from all the experiments:

\textit{Time to Verify (TTV)} of a specific observed vehicle denotes the time delay between the first proof of this vehicle being received and successful cross-verification, which involves receiving the proof from another observer in our case.
\textit{TTV} conceptually represents the time required for an observed vehicle to be covered by cross-verification.
This metric can facilitate the design of the planning module.
It can also evaluate the overall performance of the zk-PoT.

\textit{Local / Received / All Objects} represent the number of accumulated objects collected in the different modules.
Local Objects ($N_l$) is obtained from \texttt{LocalPerception};
Received Objects ($N_r$) is obtained from \texttt{CPSReceiver};
All Objects ($N_a$) is obtained from \texttt{Planner}.
These three metrics can evaluate the performance of the CPS, as well as the impact of zk-PoT in the long term.

\textit{Verification Ratio} ($R_{veri}$) is defined as the ratio of cross-verified objects ($N_a$) to the total received objects ($N_r$).
They both excluded the local perceivable objects ($N_l$) because this portion does contribute additionally to the ego vehicles.
It intuitively measures the percentage of the vehicles that could eventually be cross-verified over the long run.

\begin{equation}
    R_{veri} = \frac{N_a - N_l}{N_r - N_l}
\end{equation}

%


\textit{Bandwidth Consumption} measures the bandwidth overhead produced by the proofs, and can also evaluate the effectiveness of the bandwidth conservation of PoT3.

\subsection{Manhattan Scenario: bootstrap and convergence}
\begin{figure}[t]
    \centering
    \includegraphics[width=.7\linewidth]{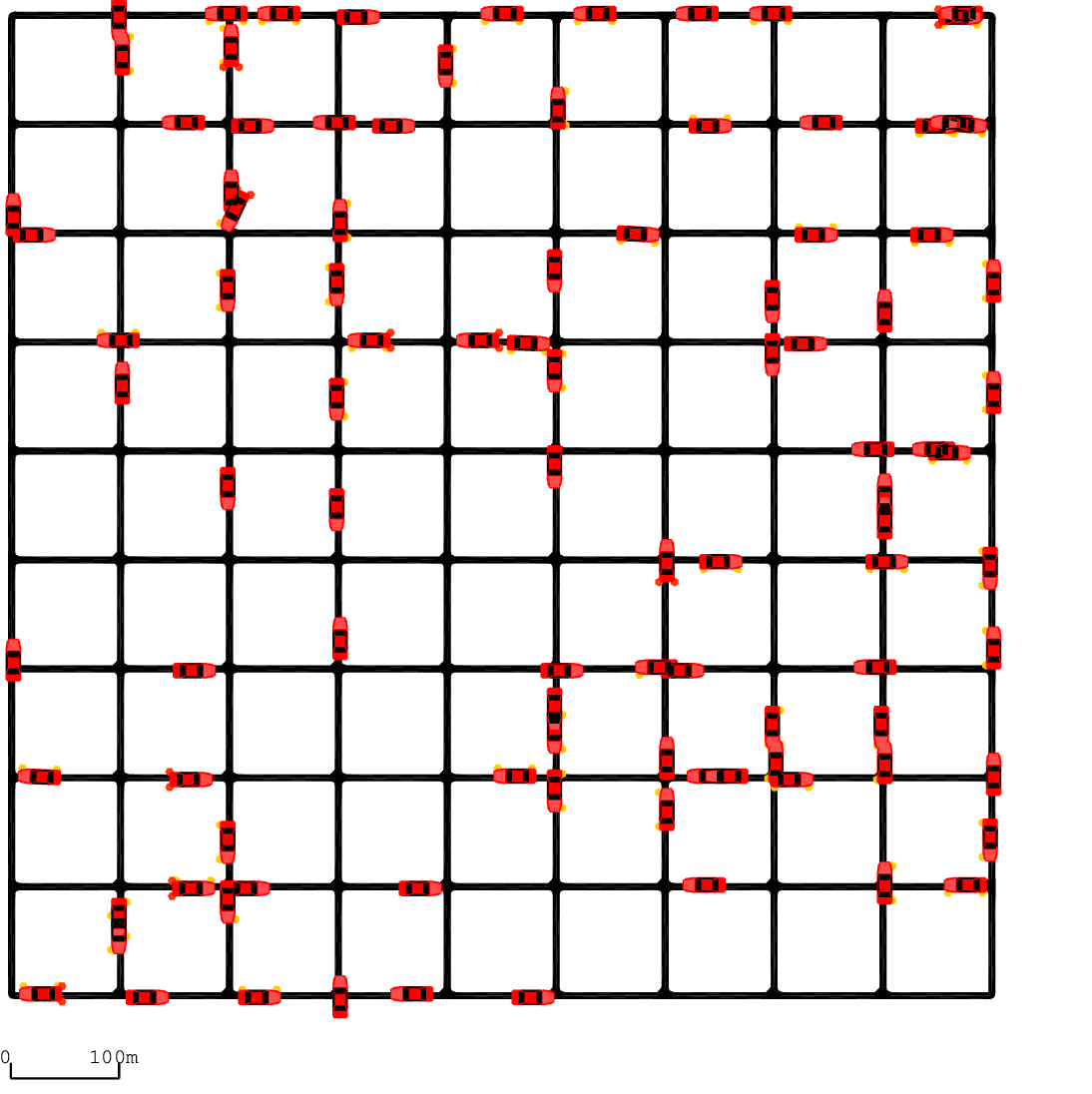}
    \caption{Manhattan scenario}
    \label{fig:sumo_manhattan_map}
\end{figure}

Firstly, a set of small-scale experiments was conducted on a synthetic grid map, also known as the Manhattan scenario.
This scenario primarily facilitates debugging and testing along with the development of Flowsim, but it can also be used to evaluate the bootstrapping and converging progress.
The map contains 10 by 10 grids with the sizes of each grid set to $100 \times 100$ meters, as shown in Fig.~\ref{fig:sumo_manhattan_map} (vehicles scaled up for better visibility).
Initially, 100 vehicles are spawned in the scenario, with random starting points and destinations.
No vehicles are spawned or removed afterward during the whole experiment.

\begin{figure}[t]
    \centering
    \includegraphics[width=\linewidth]{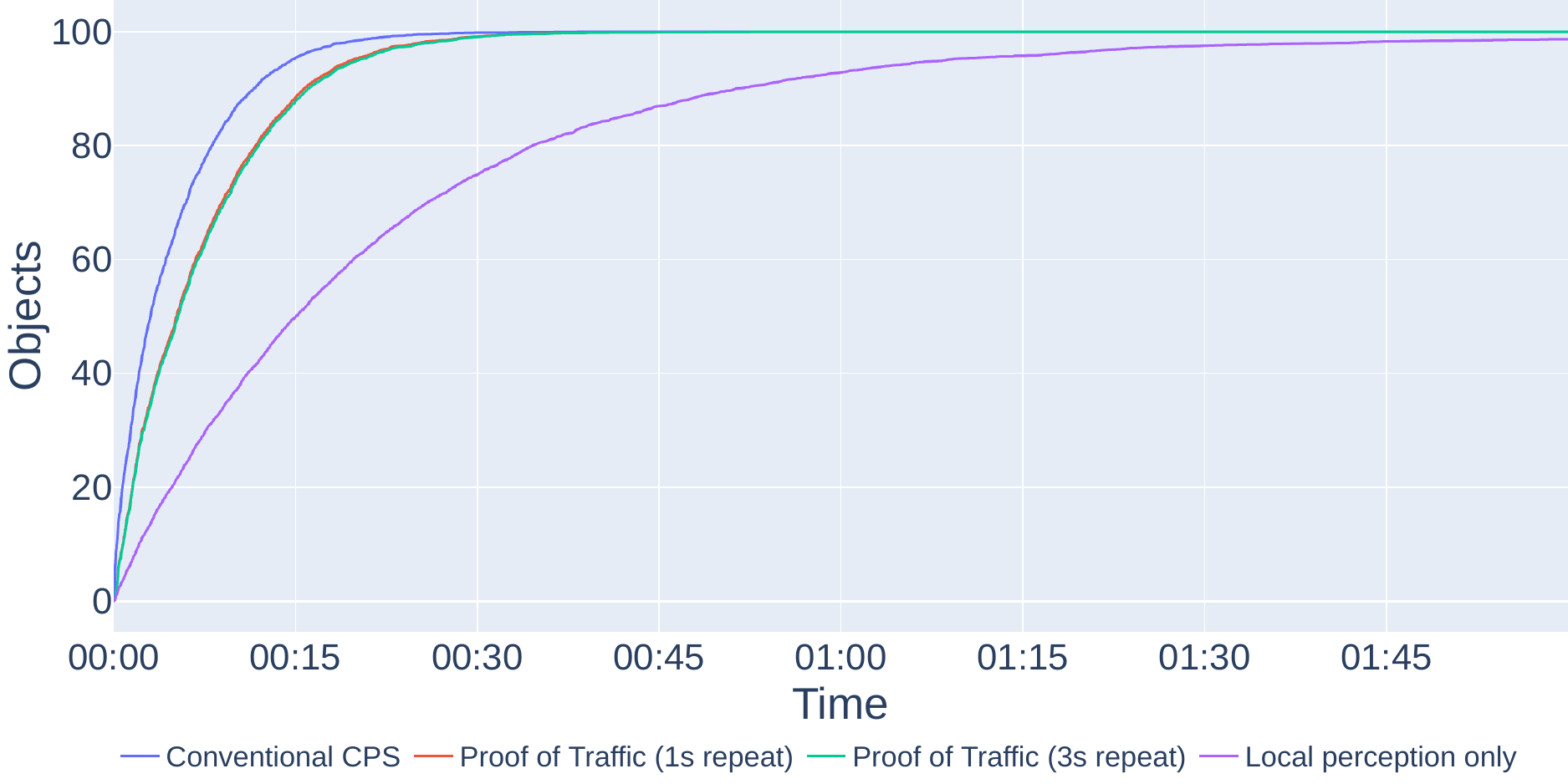}
    \caption{Accumulated available perceived objects ($N_a$) in Manhattan scenario}
    \label{fig:manhattan-objects}
\end{figure}
Each experiment was run for two hours (7200 ticks) and the number of vehicles available for \texttt{Planner} module is shown in Fig.~\ref{fig:manhattan-objects}.
\textit{Unconnected} presents the number of which no V2V communication is involved, equivalently local perception only, which converges the slowest since every vehicle needs to meet each other once.
\textit{Conventional CPS} converges the fastest since it does not require cross-verification.
\textit{Proof of Traffic (1s repeat)} converges slightly slower than the conventional non-verification approach.
\textit{Proof of Traffic (3s repeat)} converges negligibly slower than \textit{1s repeat}, but it conserves a significant amount of bandwidth, which helps in mitigating the congestion when density increases.
For comparison, the number of tracked vehicles reaches 95 percent in 14 minutes.
Conversely, Proof of Traffic reaches the same point in 20 minutes.
They both converge much faster than the unconnected setting, which takes 68 minutes.
The results demonstrate that the Proof of Traffic has a relatively fast cold start speed in such a scenario.

\begin{figure}[t!]
    \centering
    \includegraphics[width=\linewidth]{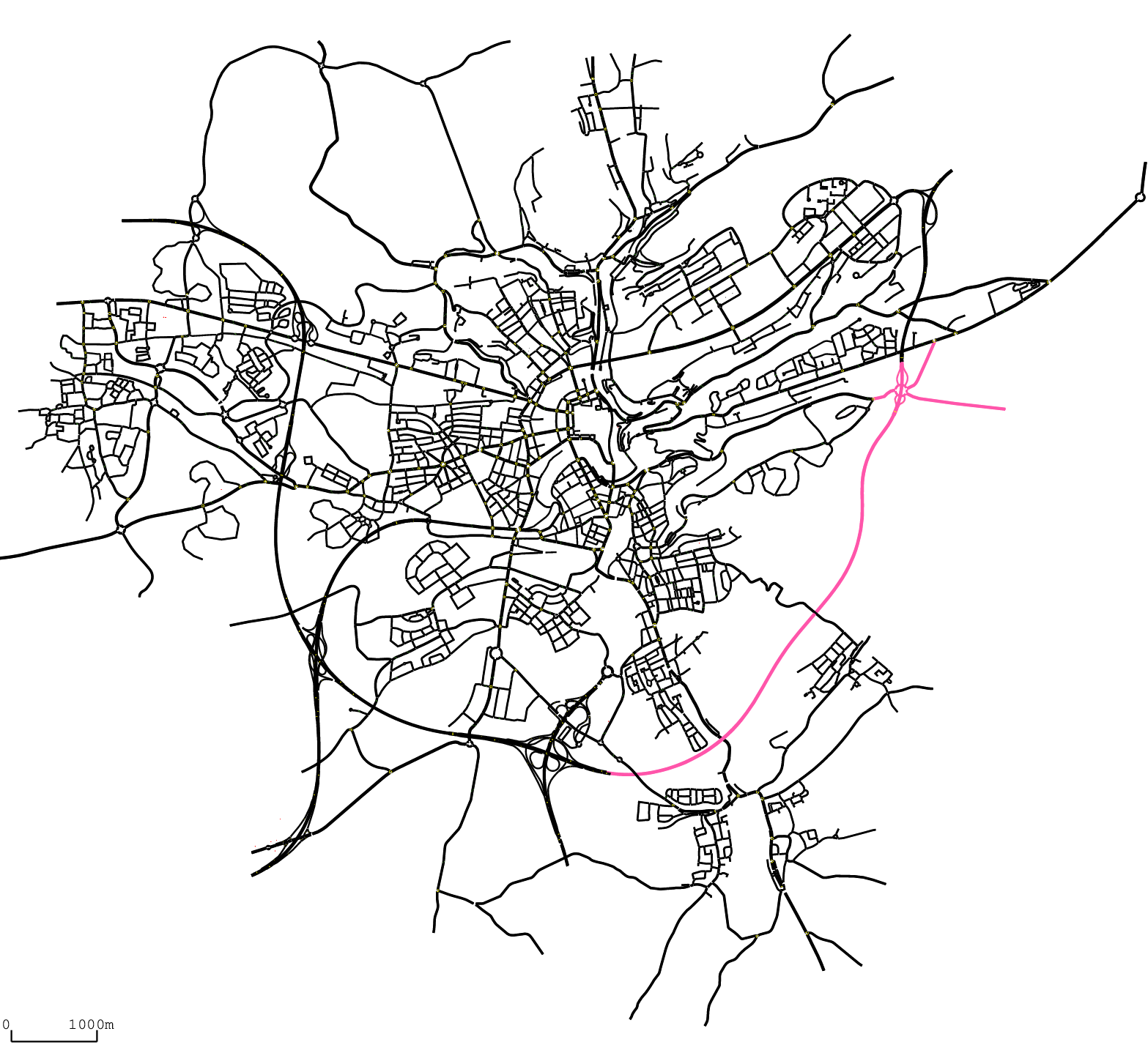}
    \caption{LuST scenario with the position of the E44 and the roundabout}
    \label{fig:sumo_lust_map}
\end{figure}

\begin{figure}[t!]
    \centering
    \includegraphics[width=\linewidth]{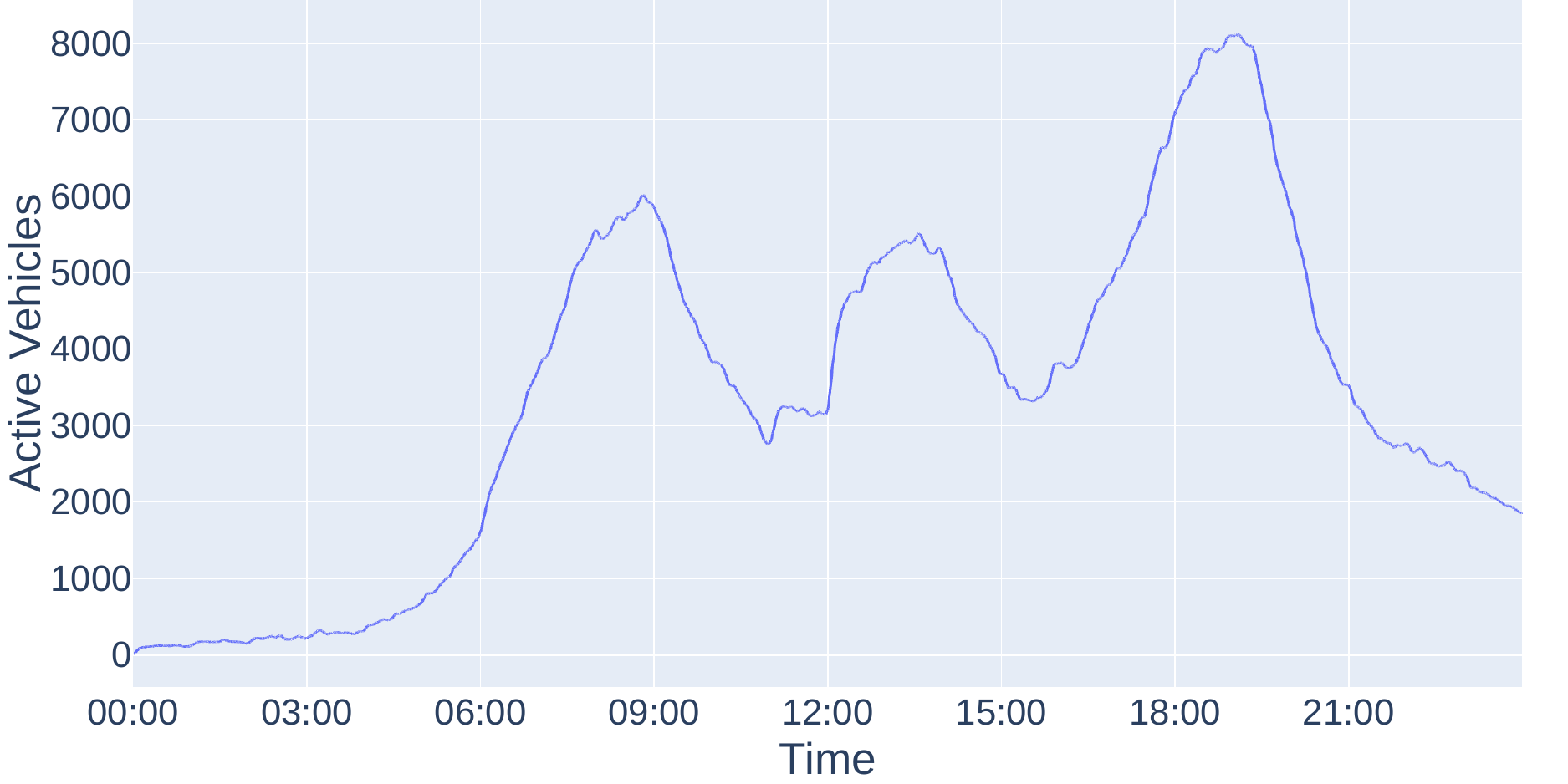}
    \caption{Number of active vehicles in the LuST scenario}
    \label{fig:lust-vehicle_count}
\end{figure}

\subsection{LuST scenario: city-scaled long-term experiments}
The final and most important group of experiments is the LuST scenario.
We use the same scenario in this work as that of the previous work, which is the 24-hour realistic traffic data generated from Luxembourg (a.k.a. the LuST scenario)~\cite{Codeca2015-wh}.
The peak number of vehicles simultaneously simulated and inspected is more than 8,000 in the evening rush hours.
\begin{figure*}[t!]
    \centering
    \includegraphics[width=\linewidth]{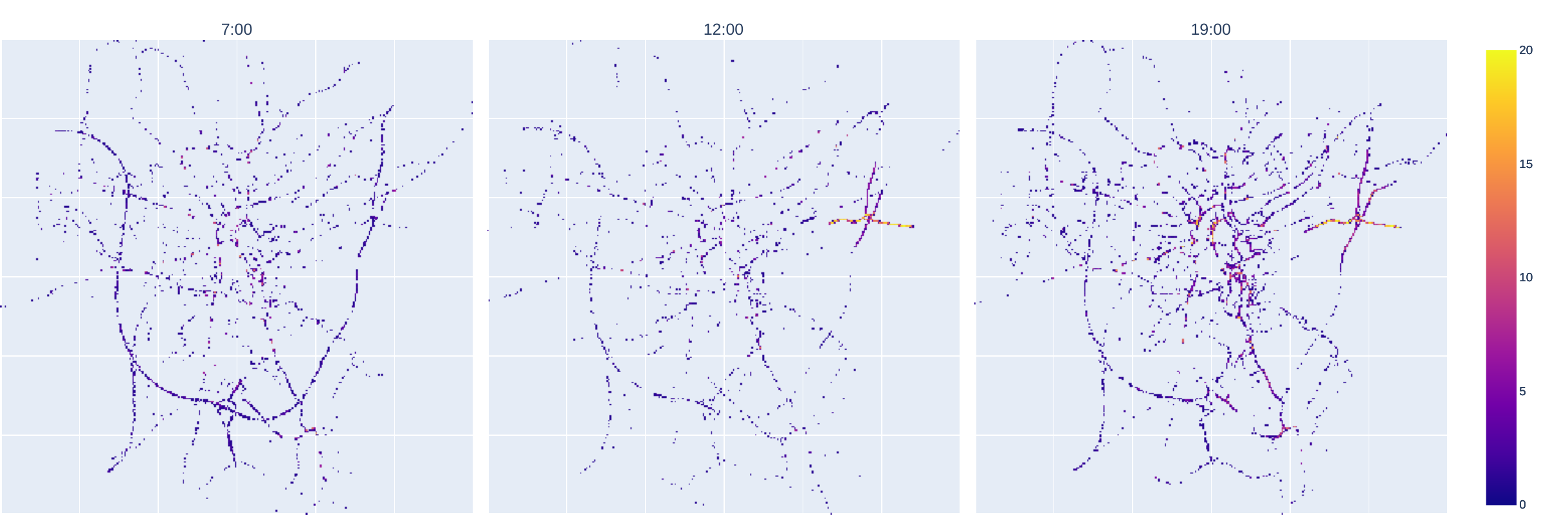}
    \caption{Number of vehicles in Luxemburg city at different times}
    \label{fig:heatmap-traffic_density}
\end{figure*}
\begin{figure*}[t!]
    \centering
    \includegraphics[width=\linewidth]{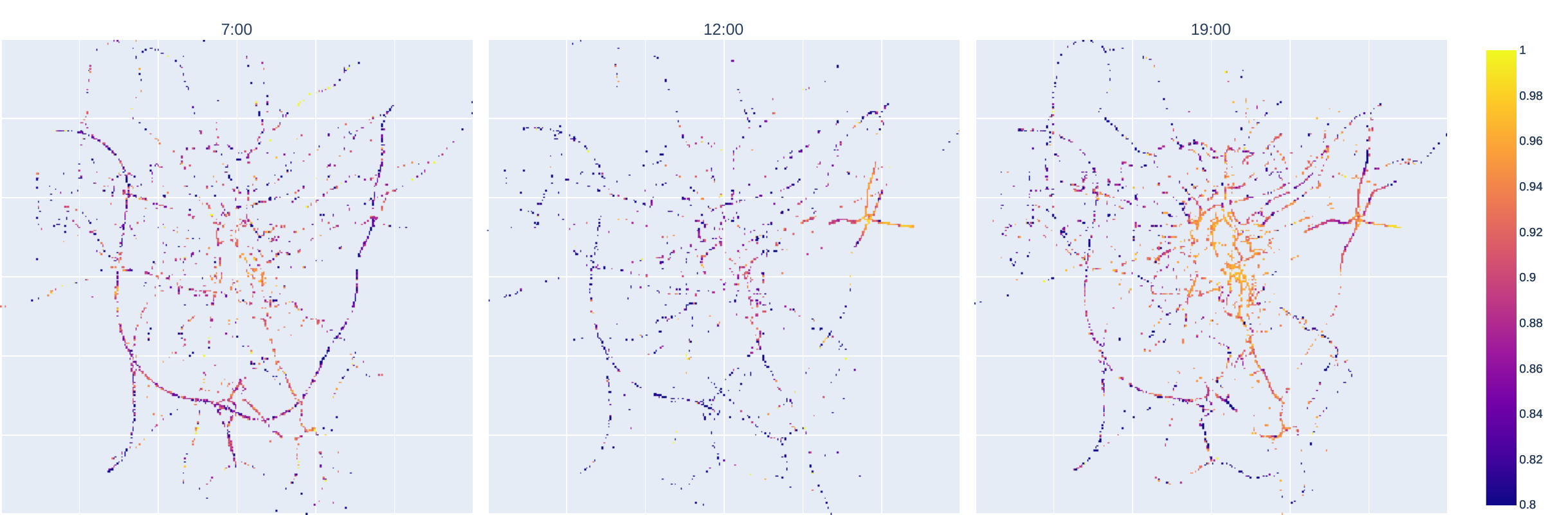}
    \caption{Verification ratio in Luxemburg city at different times}
    \label{fig:heatmap-vr}
\end{figure*}

As shown in Fig.~\ref{fig:sumo_lust_map}, the road network contains diverged types of roads and intersections, which makes it suitable to analyze the effectiveness of zk-PoT in different situations.
On the map, we can observe a U-shaped road stretching around the city.
This is the European Route E44, which is also known as the A1 Motorway of Luxembourg, the road with the most dense traffic.
However, in the LuST scenario, there is a problematic roundabout on this road, which has been highlighted in Fig.~\ref{fig:sumo_lust_map}, and will be discussed later.

We have selected the same vehicle dimensions, communication and perception parameters, and then added additional parameters for extended vehicle behaviors, such as pseudonym changing.
Flowsim is further optimized to ensure that the full-sized LuST scenario (i.e. 86400 ticks) could be finished in a wall time of one day.

\subsubsection{Traffic density and the problem}
Fig.~\ref{fig:lust-vehicle_count} depicts the number of active on-road vehicles during the day.
In the experiment, there are three rush hours in total: 10:00, 13:00, and 19:00.
Thus, we selected three samples to analyze the different behaviors of the system under different scenarios: 7:00, i.e., before the morning rush hours; 12:00, i.e., at noon; 19:00, i.e., during the peak of evening rush hours.

\begin{figure}[t!]
    \centering
    \includegraphics[width=\linewidth]{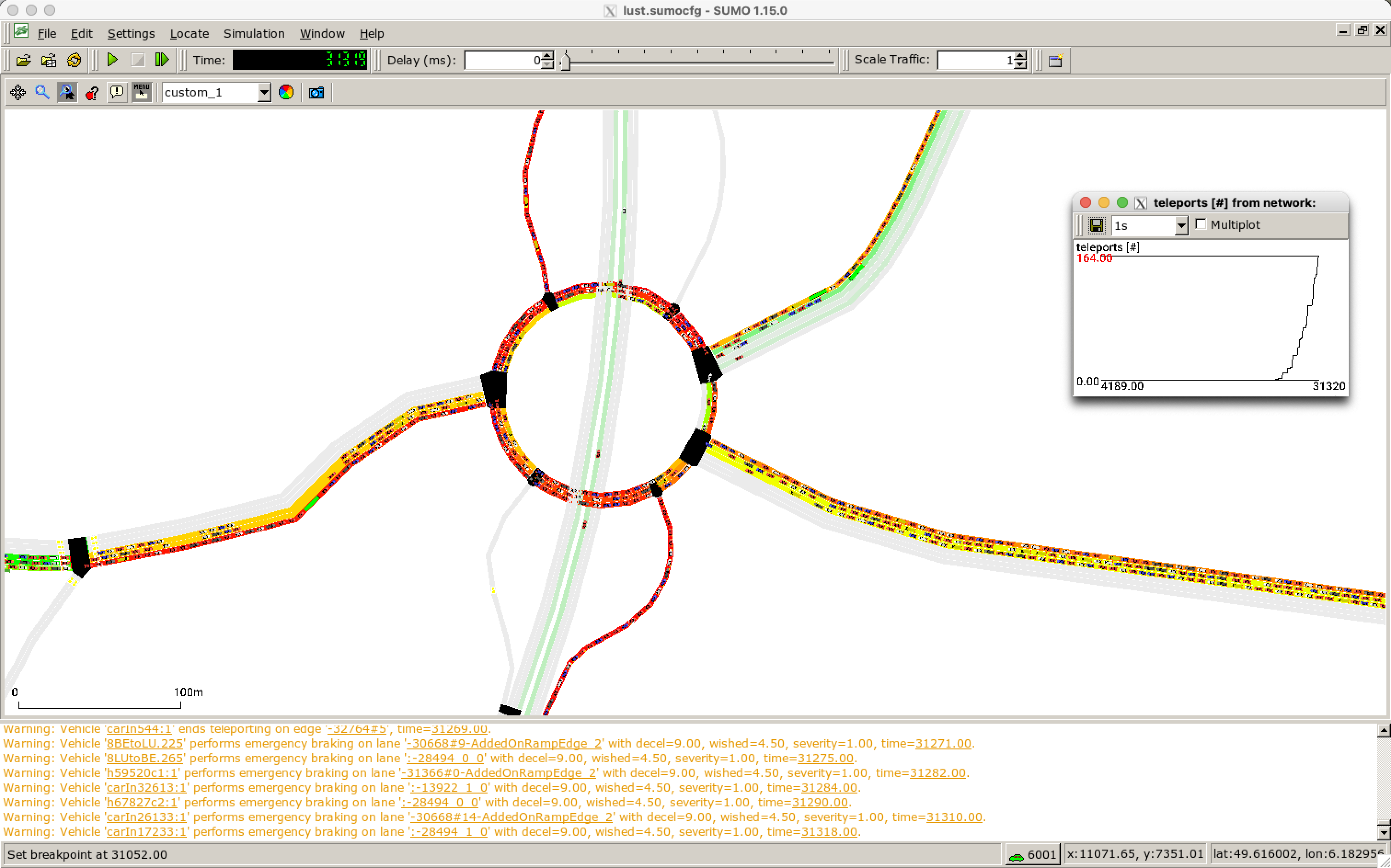}
    \caption{Vehicles and lane occupancy near the E44 roundabout at 08:42}
    \label{fig:31319}
\end{figure}

Fig.~\ref{fig:heatmap-traffic_density} depicts the traffic density distribution in the three samples.
In the heatmap of 7:00, the data points form a notable U-shaped path, representing the normal traffic on the E44.
However, in the sub-figure of 12:00 and 19:00, we can see the data points on this road are very sparse, particularly in the southeast region, i.e., the bottom right area.
This scarcity is caused by the complete congestion observed at the roundabout shown in Fig.~\ref{fig:31319}; refer to the marked part in Fig.~\ref{fig:sumo_lust_map}.
In Fig.~\ref{fig:31319}, the color of the lane represents their occupancy.
Notably, the occupancy at E44 appears to be very low, while the roundabout and the entrances/exits of E44 are completely blocked by vehicles.
This severe congestion has even produced a large number of teleportation events~\cite{Documentation_undated-tr}, where vehicles approaching the roundabout from other roads were instantaneously teleported to different locations.
The congestion prevents vehicles from entering E44, leading to missing data in the southeast corner of the road during those time intervals in Fig.~\ref{fig:heatmap-traffic_density} and~\ref{fig:heatmap-vr}.

We believe that this is caused by the well-known border traffic issue and the teleportation issue of SUMO scenarios; however, we have not modified the LuST scenario itself to ensure that it corresponds with that of other researchers.
Nevertheless, we can still analyze the other part of E44 to observe behavior in similar highway traffic situations.

\subsubsection{Verification ratio}
The verification ratio is the main performance metric of the system.
We measure the long-term verification ratio throughout the entire day, which concurs more with the real-world conditions, instead of calculating the verification ratio (named as success ratio in the last work) separately for each tick like in the last work.

Fig.~\ref{fig:heatmap-vr} depicts the verification ratio in the same sampling time.
We can observe that the verification ratio has a positive correlation to vehicle density.
It starts at approximately 80~\% at 7:00 and peaks at over 96~\% at 19:00.
Furthermore, the vehicles in the downtown area generally perform better than the ones on the highway.
This is because the vehicles on the highway are typically platooning and did not encounter different vehicles for a relatively long period.
Conversely, the diversity of roads in the downtown area and the randomness of the vehicle behaviors in that area make the platoons of vehicles very volatile, thus increasing the cooperative perception and the verification ratio.
This phenomenon is particularly evident in the figure indicating the scene of 19:00.


\subsubsection{Time-to-Verification}
\begin{figure}
    \centering
    \includegraphics[width=\linewidth]{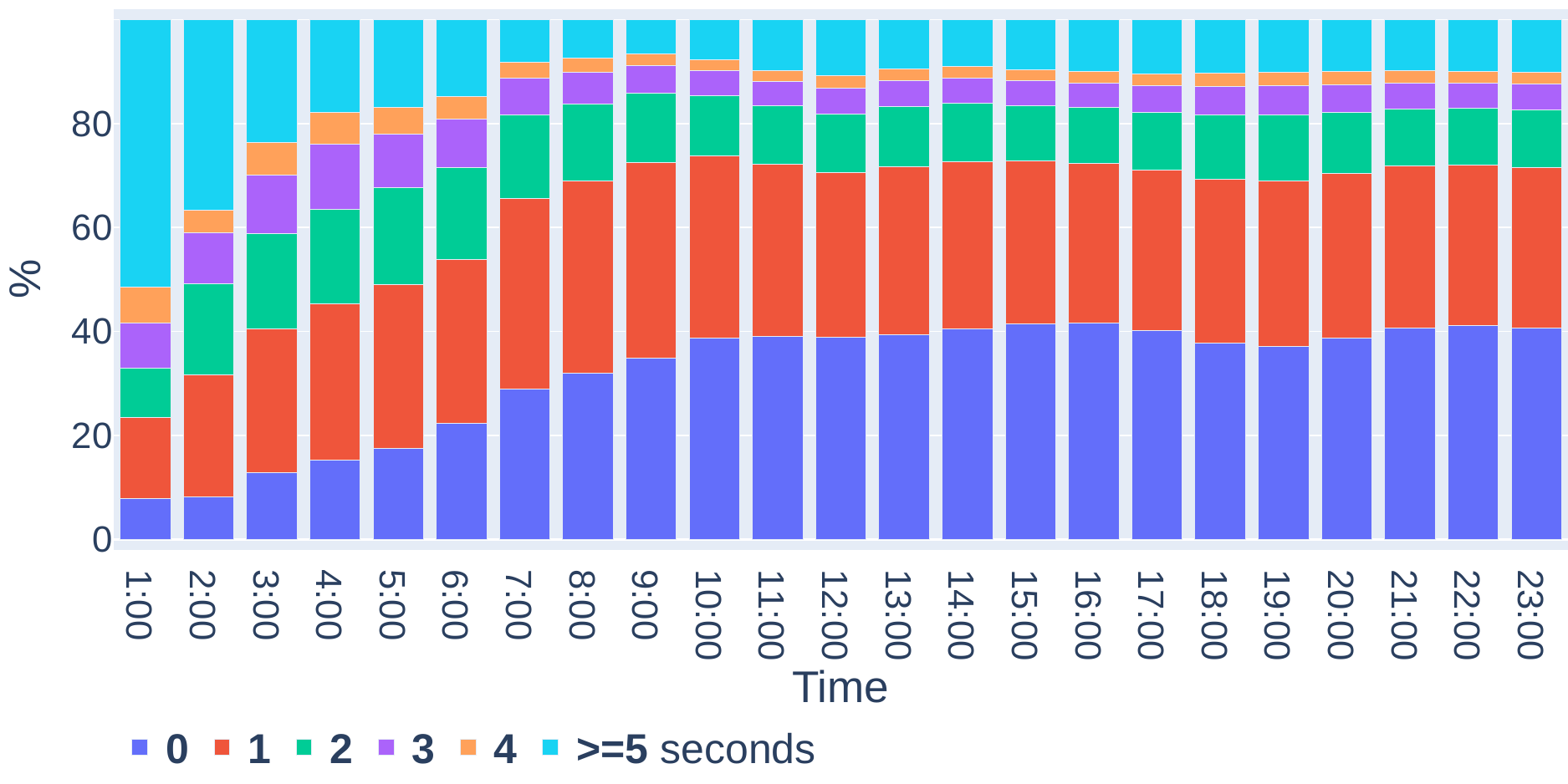}
    \caption{Time-to-Verification (TTV) distribution}
    \label{fig:ttv}
\end{figure}

The timeliness of the cooperative perception data, i.e. the TTV, is the second thing to be considered.
Fig.~\ref{fig:ttv} depicts the TTV distribution of every hour, scaled to 100~\%.
We can observe that the system is bootstrapping, and that the traffic is sparse during the period from 0:00 to 7:00, resulting in longer verification delays.
From the data after 9:00, we can observe that the trend becomes relatively flat, indicating that the system has reached dynamic equilibrium. 
Based on the trend, we can observe that approximately 35~\% of the data can be verified in the same tick that it first appears, which produces a sub-second verification delay.
Furthermore, over 65~\% and 80~\% of the observations could be verified within 1 s and 2 s, respectively. 
The result indicates that the proposed zk-PoT presents a very short latency from the point of receiving data to considering them trustworthy.
Conversely, the conventional statistics and plausibility approaches require a significantly longer time to collect ``evidence,'' to build the trust of either the sender or the received data.

\subsubsection{Protocol overhead}
\begin{figure}[t!]
    \centering
    \includegraphics[width=\linewidth]{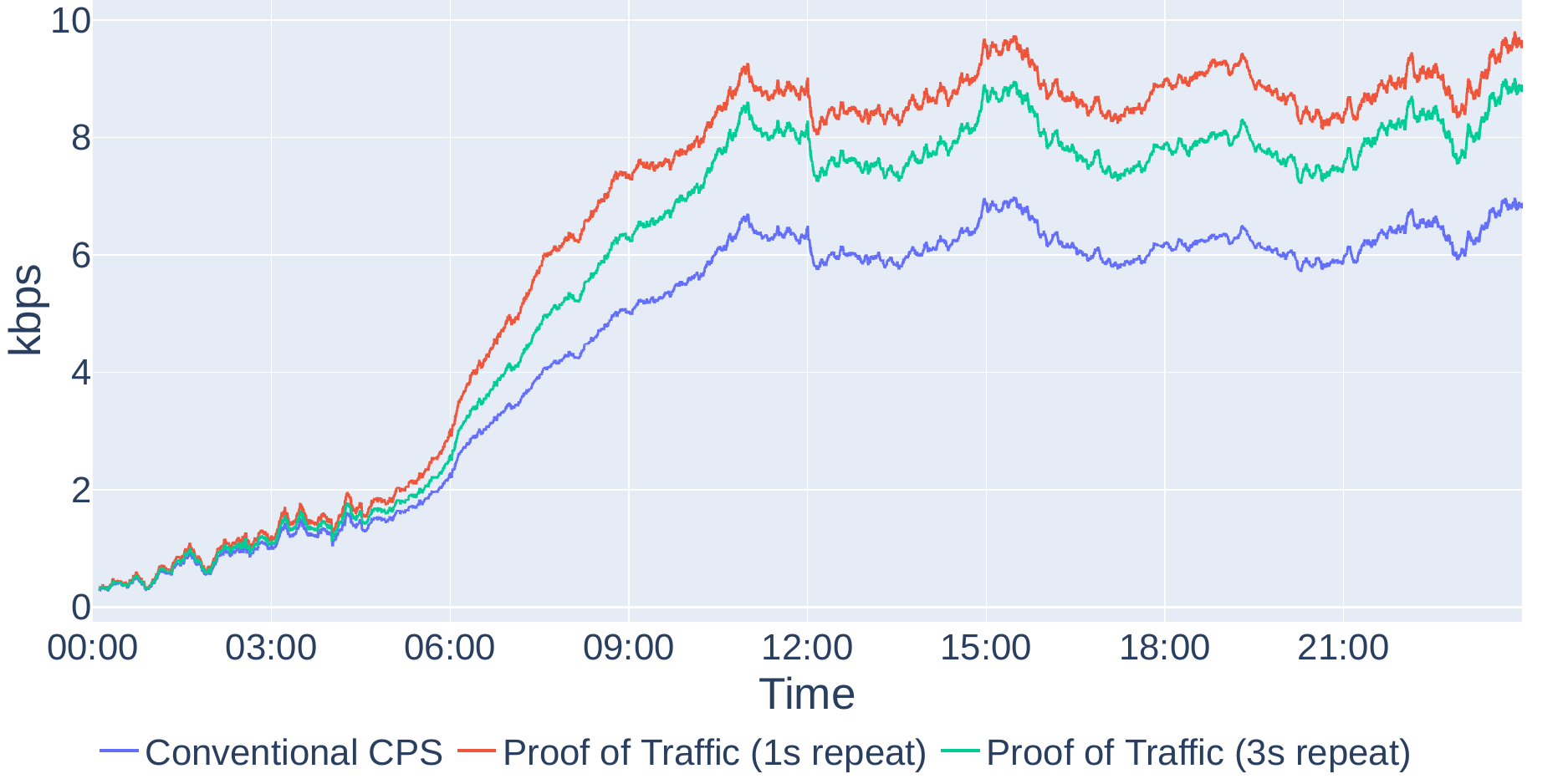}
    \caption{Bandwidth consumption per vehicle}
    \label{fig:line-bandwidth-consumption}
\end{figure}

Lastly, we must consider the protocol overhead as well.
Fig.~\ref{fig:line-bandwidth-consumption} shows that the system is not fully bootstrapped in the simulation time before 9:00, similar to the previous pattern.
Beyond this point, we can consider it as the normal working condition of our system.
Therefore, we primarily focus on analyzing the differences in this phase.

In the conventional CPS with no PoT enabled, the average bandwidth consumption of a single vehicle stabilizes at approximately 6 kbps.
In the PoT-enabled system with proofs repeated every second, the system presents a 40~\% increase in bandwidth when compared with the baseline system, reaching 9 kbps.
However, if we change the PoT to repeat proofs every 3 s, the bandwidth increment reduces to about 25~\%, resulting in a bandwidth of 7.5 kbps, which is considered stable and feasible.
Furthermore, the bandwidth did not fluctuate according to the total number of vehicles, which is attributed to the dynamic equilibrium and saturation effects.

\subsection{Comparison with other approaches}

\begin{table*}[t!]
    \centering
    \begin{tabular}{lllll}
    \hline
    Method & \cite{Zhang2022-pe} & \cite{Ambrosin2019-ic} & MISO-V~\cite{Liu2021-lg} & \textbf{zk-PoT} \\
    \hline
    Type & MBD/Node/Trust & MBD/Data/Plausibility & MBD/Data/Consistency & \textbf{Prove based}\\
    Applicable attack types & Omission & Ghost & (Similar to) Ghost/Omission & \textbf{Ghost} \\
    Evaluation method & Simulation/T-Junction & None (Framework) & Simulation/T-Junction & \textbf{Simulation/City-scale}\\
    Verification ratio & N/A & N/A & 80\% & \textbf{90-95\%} \\
    False positive ratio & N/A & N/A & 20\% & \textbf{0} \\
    Decision delay & 12.5 sec & N/A & 0 sec & \textbf{>80\% in 2 sec}\\
    \hline
    \end{tabular}
    \caption{Comparison of zk-PoT with misbehavior detection methods}
    \label{tab:comparison}
\end{table*}

Drawing upon the quantitative results presented earlier, a comparative analysis with conventional misbehavior detection approaches can be conducted.
While MBDs assume data are trustworthy and try to detect lies, we assume the opposite and try to prove truths.
Despite the different intentions and methodologies employed by these two methods, they converge in providing similar functionality, specifically ensuring data authenticity.

Table~\ref{tab:comparison} presents a comparative analysis between zk-PoT and various MBD approaches, specifically those focusing on cooperative perception.
The first classification is methodological taxonomy, distinguishing between node-centric and data-centric approaches.
Within the data-centric domain, two representative strategies, namely plausibility-based and consistency-based methods, were selected for examination.
The second categorization involves the identification of applicable attack types, specifically ghost vehicle attacks (fabricating non-existent vehicles) and omission attacks (intentionally excluding vehicles).

The analysis reveals that zk-PoT demonstrates notable attributes, including a zero false positive ratio, an elevated verification ratio, and a comparatively expedient decision delay.
However, it is imperative to acknowledge zk-PoT's inherent limitation, notably its susceptibility to omission attacks.
This limitation is inherent to zk-PoT's fundamental nature:
non-existent means no data and thus no entropy at all, so it's impossible to prove that ``nothing is there''.

\subsection{Adaptability to non-V2X vehicles}
In the zk-PoT mechanism, an imperative requirement for the identification and subsequent verification of a target vehicle have a unique identifier with sufficient entropy.
For the sake of simplicity, this study assumes the utilization of the V2X station ID of each vehicle.
It is noteworthy, however, that such an arrangement is not obligatory.

Different technologies already in the market also feature their distinct identifiers, thereby offering flexibility in implementation and deployment.
Notably, regions such as certain states in the United States exhibit a notable prevalence of Radio-Frequency Identification (RFID) based tags~\cite{Rich2008-xq,Shbaklo2016-up}, while Japan witnessed an escalating adoption of Electronic Toll Collection Systems 2.0 (ETC 2.0)~\cite{Makino2015-zk}.
The vehicles equipped with those existing technologies can thus still be observed and proved, even without V2X capabilities itself.
zk-PoT can be adapted to utilize some or all of those technologies with little effort.

\section{Threat Analysis}
\label{sec:analysis}
The zk-PoT provides vehicles with the capability to swiftly and deterministically validate data received from the CPS. 
This eliminates the possibility of na\"ive attacks and data tampering, such as fabricating non-existent vehicles and duplicating existing vehicles.
Furthermore, various complex attacks can be prevented or mitigated.
Although this list is not exhaustive, it covers some of the most common attacks that can be encountered.

\subsection{Brute-Force Attack and Dictionary Attack}
Malicious entities may attempt to perform brute-force attacks or dictionary attacks on the number plate used in the proof.

In the case of a brute-force attack, the attacker can try to match the public keys used in other vehicles' proofs by only ``hearing'' the candidate observed vehicles.
As the range of ``hearing'' is considerably larger than that of ``seeing,'' attackers can still pretend that they ``saw'' the target vehicles.
Furthermore, this attack can be done in an offline fashion and does not require any exchange of information with other vehicles.
We can eliminate these types of attacks by employing key derivation functions (KDFs), such as PBKDF2~\cite{Kaliski2017-gr} or scrypt~\cite{Percival2016-px}, to increase the length of time required for a successful brute-force attack, making it longer than the lifespan of a pseudonym.

In the case of a dictionary attack, the attacker can gather previously known number plates and use them as a reference.
This type of attack presents a higher success rate than that of a brute-force attack due to the locality of the vehicles.
Mechanics like ephemeral salt can be employed to prevent this attack.

\subsection{Location Privacy Against Tracking}
The use of pseudonyms is crucial in ensuring location privacy in V2X communications.
With the combination of pseudonyms and KDFs involved, brute-force attacks cannot be performed to create a timely proof.
However, an attacker can still collect data on the road and recover a vehicle's trajectory.

Fortunately, after the target changes its pseudonymous ID, remote attackers can no longer track it, even if they have information about its old ID and number plate.
As such, the proposed mechanism provides at least the same level of privacy as that of traditional pseudonyms.

\subsection{Spam Attack}
Verifiers are vulnerable to spam attacks as partial proofs cannot be falsified,
Attackers can easily create seemingly valid proofs randomly, 
which can deplete the victim verifiers' computing power and memory, thus potentially causing a denial-of-service situation.

This type of attack can be mitigated by setting a limit on the number of unmatched proofs from each prover.
If a vehicle exceeds this limit, the verifier must disregard any subsequent proofs sent by the vehicle until some of them are matched by other vehicles.

\subsection{Signal Jamming Attack}
In the proposed scheme, a malicious vehicle cannot simply replay a message.
However, the attacker can still jam the signal from the original prover and then replay its proof as their own.

This kind of attack can be mitigated by including the prover's own pseudonym in the proof.
This is because the pseudonym of a vehicle cannot be easily forged, making it a secure method of identifying the original prover.

\subsection{Collusion Attack and Sybil Attack}
Collusion and Sybil attacks are both types of attacks that are specific to the applications that require voting or cross-verification, including the proposed method.
A collusion attack is a type of attack in which multiple malicious entities work together to deceive a system.
In a Sybil attack~\cite{Douceur2002-vf}, the attacker creates multiple pseudonymous identities to impersonate multiple users.
These two types of attacks can be combined to create a set of fake vehicles that can deceive victim verifiers.

While these attacks are typically challenging, we can implement certain strategies to mitigate them.
We can mitigate the Sybil attack by implementing mechanisms that prevent the simultaneous use of different pseudonyms belonging to the same vehicle.
The design of pseudonyms is not directly related to the proof system and will be addressed in the future.
For the non-Sybil collusion attack, the number of vehicles controlled by the attacker is limited as they require a fleet to perform the attack.
In this case, we can limit the revenue from the data.
The revenue could comprise a trust value or credits, and it varies depending on the system built based on the proof system.
For example, we can set the trust value for a successful match between two specific provers to be the reciprocal of the number of accepted matches over the past 15 minutes.
This countermeasure can cause a sharp decrease in trust, making further collusion unprofitable.

\section{Conclusion}
\label{sec:conclusion}
This study proposed a novel deterministic cross-verification scheme called zk-PoT.
By enabling vehicles to generate zero-knowledge blind proofs to their independent observation, the remote parties can be convinced by cross-verifying these proofs, without leveraging the ground truth or trust evaluation.
Subjecting to the zero-knowledge property, the zk-PoT will not reveal any information about any particular vehicle, thereby preserving the location privacy of the vehicles.

The quantitative simulation analyses in the Luxemburg scenario verify that zk-PoT is in good performance.
In the experiments, about 80~\% to 96~\% of observed vehicles can be cross-verified by zk-PoT.
Over 80~\% of cross-verification happened within a sub-2 s delay and over 90\% happened within 5 s.
The bandwidth consumption overhead is approximately 25~\% compared to the original 1 Hz CPS standard.

The threat analysis against various kinds of attackers shows that zk-PoT could survive most of the common threats, such as brute-force attacks, dictionary attacks, spam attacks, and signal jamming attacks.
It can maintain location privacy to the same level as the original pseudonym system.
The collusion attack is challenging, but it can still be mitigated by extra rules.

Zk-PoT can be either used as a standalone method or integrated with existing cooperative perception standards, as shown in the aforementioned example, where zk-PoT is implemented for the CPS in the ETSI/ITS standards.
Furthermore, zk-PoT is particularly helpful for bootstrapping the trust establishment process in several existing trust management models as they can provide strong evidence that can be cross-verified by other vehicles~\cite{Hussain2021-ki}.

Owing to the limited scope of this study, not all aspects have been examined comprehensively.
The following issues should be considered in the future.
Firstly, the vehicles are not incentivized to share data and tend to be selfish.
This will reduce the number of proofs thus harming the cross-verification ratio.
Based on the ``privacy-trust dilemma,'' an economic model could be proposed.
Secondly, Sybil attacks are not solved even when collusion mitigation techniques are introduced.
This is a potential research direction to enhance zk-PoT.
Lastly, the non-connected vehicles are not considered in this study because they lack identity with enough entropy.
Some other mechanics should be considered to make zk-PoT better adapted to mixed traffic environments.

%

\bibliographystyle{unsrt}
\bibliography{main,survey}

\begin{IEEEbiography}[{\includegraphics[width=1in,height=1.25in,clip,keepaspectratio]{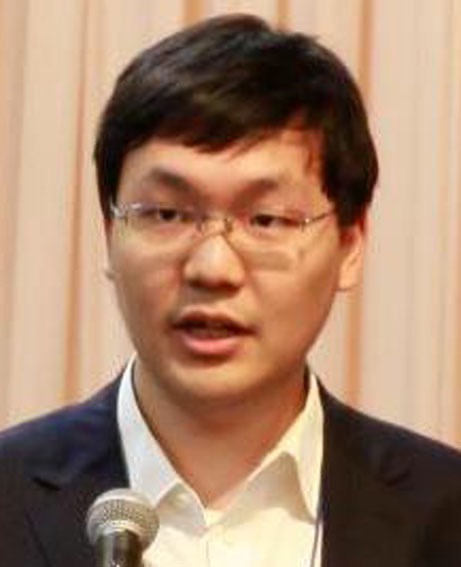}}]{Ye Tao} received the B.S. degrees in Beijing University of Posts and Telecommunications in 2014 and M.S. degree in The University of Tokyo in 2016.
He is currently pursuing his PhD at the Graduate School of Information Science and Technology, at the University of Tokyo.
His current research interests are cybersecurity, cryptography, and networking for connected autonomous vehicles.
\end{IEEEbiography}

\begin{IEEEbiography}[{\includegraphics[width=1in,height=1.25in,clip,keepaspectratio]{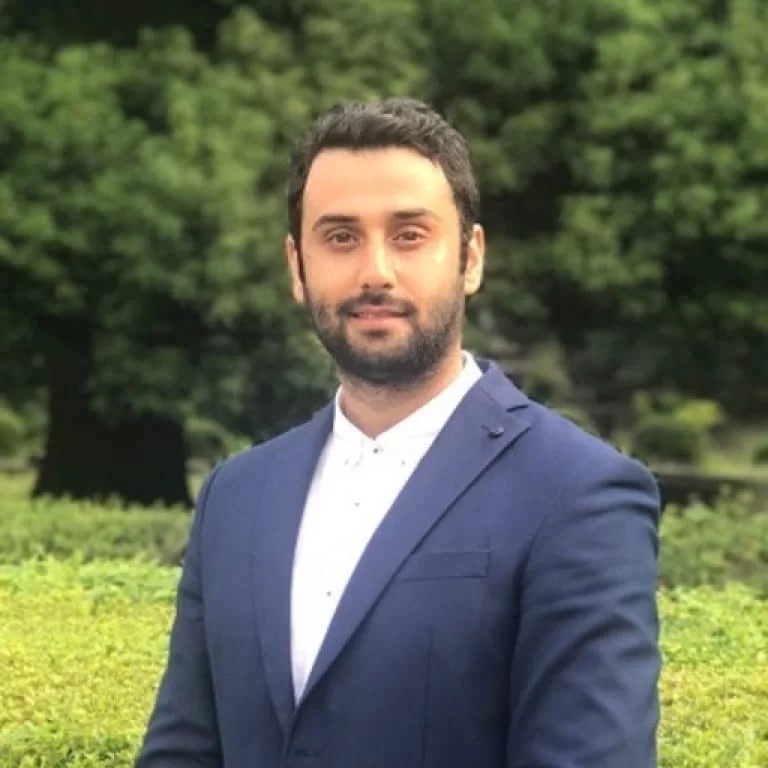}}]
{Dr. Ehsan Javanmardii}
received his M.E. degree in computer architecture from the Amirkabir University of Technology, Iran, in 2012 and a Ph.D. degree in information and communication engineering from The University of Tokyo, Japan, in 2018. From 2014 to 2015, he was a Visiting Research Student at The University of Tokyo. He was also a Visiting Student Researcher at the University of California, Berkeley, from 2016 to 2017. He also completed the Graduate Program for Social ICT Global Creative Leaders with The University of Tokyo in 2018. He has been a Post-Doctoral Researcher with the Institute of Industrial Science, The University of Tokyo since 2018. He is a project assistant professor at the Graduate School of Information Science and Technology, the University of Tokyo, Japan. His research interests include intelligent vehicles, autonomous vehicles’ self-localization, mobile mapping systems, ADAS maps, mapping, sensor fusion, and vehicle perception. He received the IEEE Intelligent Transportation Systems Society Best Student Paper Award in 2017.
\end{IEEEbiography}

\begin{IEEEbiography}[{\includegraphics[width=1in,height=1.25in]{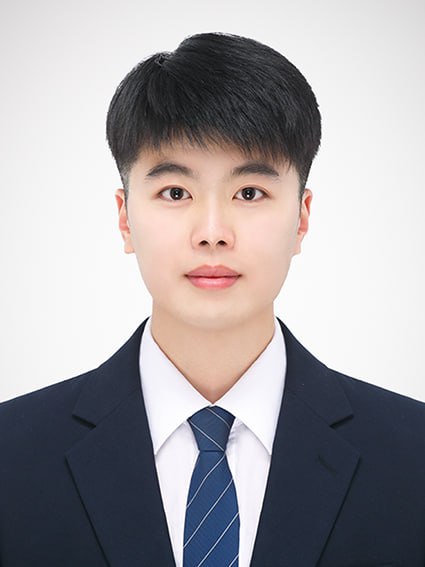}}]{Pengfei Lin}
received the B.S. degree in electrical engineering and automation from Northwestern Polytechnical University in 2018 and M.S. in electrical engineering from Hanyang University in 2021. He is currently pursuing his Ph.D. in creative informatics from The University of Tokyo. 
His current research interests include path planning, collision avoidance, and model predictive control for autonomous vehicles. He was an invited reviewer for the \textit{\textsc{IEEE Robotics and Automation Letters}}, \textit{\textsc{IEEE International Conference on Intelligent Transportation Systems}}.
\end{IEEEbiography}

\begin{IEEEbiography}[{\includegraphics[width=1in,height=1.25in,clip,keepaspectratio]{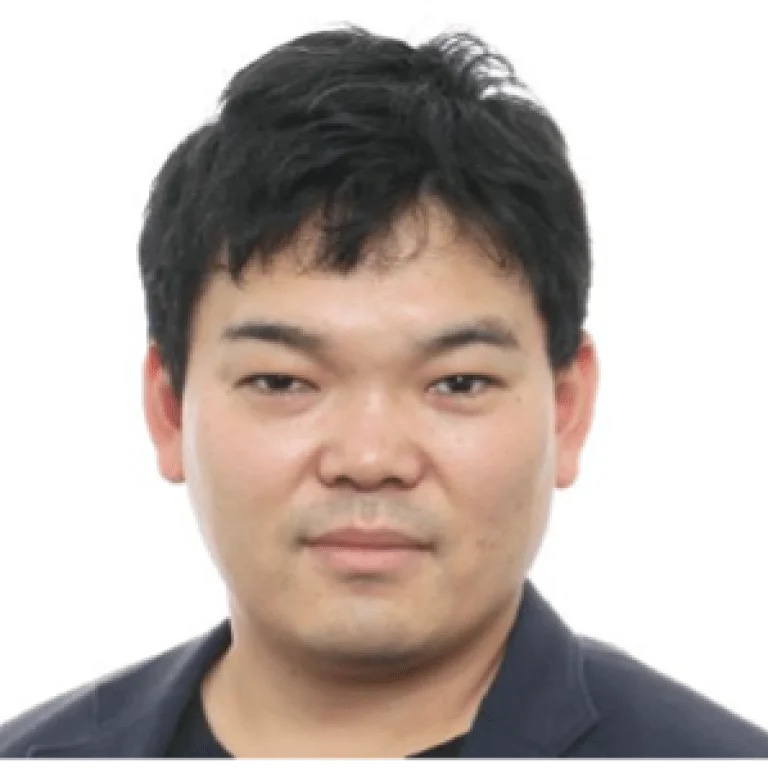}}]
{Dr. Jin Nakazato}
(Member, IEEE) is currently a Specially Appointed Assistant Professor at the University of Tokyo. He also works as a Postdoctoral Researcher with the Tokyo Institute of Technology. He received B.E. and M.E. degrees from the University of Electro-Communications, Japan, in 2014 and 2016, respectively. He received a Ph.D. degree from the Tokyo Institute of Technology, Japan, in 2022. From 2016 to 2020, he was with FUJITSU LIMITED. From 2020 to 2022, he was with Rakuten Mobile, Inc. His research interests include Multi-access Edge Computing, NFV/SDN Orchestrator, V2X, Open RAN, UAV networks, and virtualization RAN. He is a Member of IEICE, IEEE, and WIDE Project. He received the Best Paper Award of the 11th International Conference on Ubiquitous and Future Networks (ICUFN 2019) in 2019. He serves as an editor of IEICE Communications Express (ComEX) is a peer-reviewed open access letter journal covering the entire field of communication.

\end{IEEEbiography}

\begin{IEEEbiography}[{\includegraphics[width=1in,height=1.25in,clip,keepaspectratio]{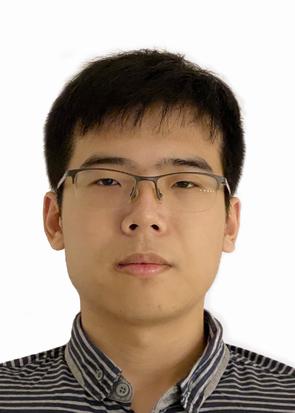}}]
{Yuze Jiang} earned his B.S. degree in Computer Science from the University of Minnesota in 2020. He is presently a Master's candidate in the Department of Creative Informatics at the University of Tokyo. His research interests encompass a range of topics within Intelligent Transportation Systems, primarily concentrating on V2X communications and vehicle cooperative localization techniques.
\end{IEEEbiography}

\begin{IEEEbiography}[{\includegraphics[width=1in,height=1.25in,clip,keepaspectratio]{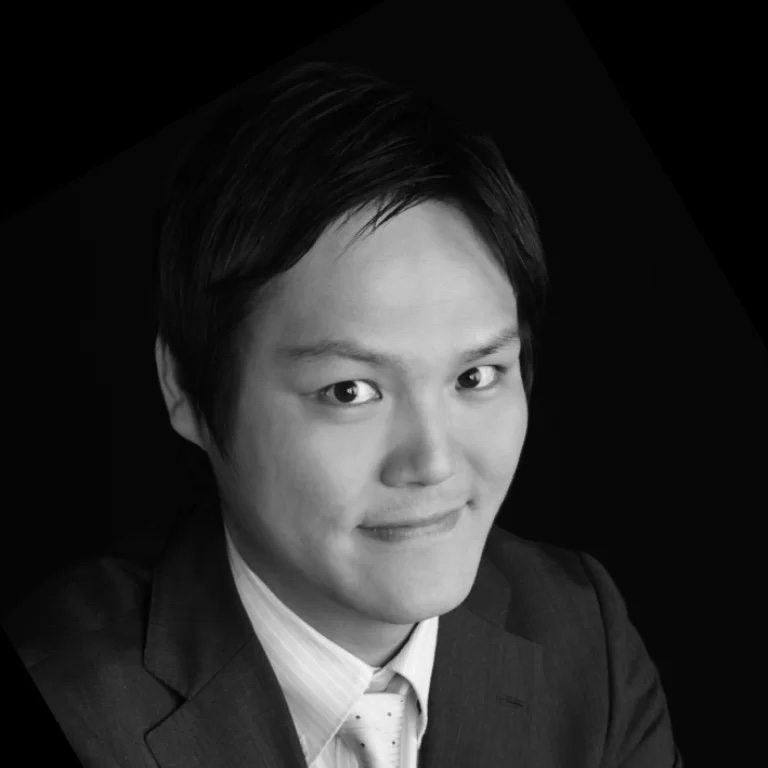}}]
{Dr. Manabu Tsukada}
is currently an associate professor at the Graduate School of Information Science and Technology, the University of Tokyo, Japan. He is also a designated associate professor at the Center for Embedded Computing Systems at Nagoya University, Japan.  He was a visiting professor at Aalto University from February 2021 to November 2021. He received his B.S. and M.S. degrees from Keio University, Japan, in 2005 and 2007, respectively. He worked in IMARA Team Inria, France, during his Ph.D. course and obtained his Ph.D. degree from Centre de Robotique, Mines ParisTech, France, in 2011. During his pre and postdoc research stages, he has participated in a multitude of international projects in the networked ITS area, such as GeoNet, ITSSv6, SCORE@F, CVIS, Nautilus6, and ANEMONE. He served as a board member of the WIDE Project 2014-2022. His research interests are mobility support for the next-generation Internet (IPv6), Internet audio-visual media, and communications for intelligent vehicles.
\end{IEEEbiography}

\begin{IEEEbiography}[{\includegraphics[width=1in,height=1.25in,clip,keepaspectratio]{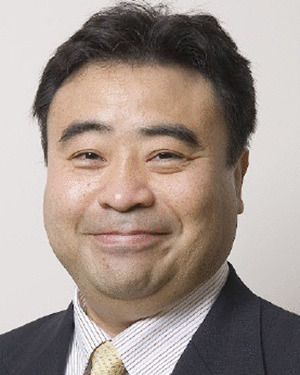}}]
{Dr. Hiroshi Esaki} (Member, IEEE) is currently a Professor with the Graduate School of Information Science and Technology, University of Tokyo, Bunkyo, Japan. In 1987, he joined Research and Development Center, Toshiba Corporation, where he engaged in the research of ATM systems. From 1990 to 1991, he has been with Bellcore Inc., New Jersey, USA, as a residential Researcher. From 1994 to 1996, he has been with Columbia University, New York, NY, USA. He has proposed the CSR architecture that is one of the origin of MPLS (Multi-Protocol Label Switching), to the IETF and to the ATM Forum. From 1998, he was a Professor with the University of Tokyo, and a Board Member of WIDE Project. From 1997, he has involved in many of the IPv6 research and development at the WIDE project. He is a cofounder of series of IPv6 special project in the WIDE project, (KAME Project, TAHI Project USAGI Project). He is an Executive Director of IPv6 promotion council, the Vice Chair of JPNIC (Japan Network Information Center), and the Chair of IPv6 Ready Logo Program run by IPv6 Forum.
\end{IEEEbiography}

\EOD

\end{document}